\documentclass[useAMS,usenatbib]{mn2e}

\title[GMC formation by agglomeration and self gravity]
{GMC formation by agglomeration and self gravity}
\author[C. L. Dobbs]
{C. L. Dobbs\thanks{E-mail:
dobbs@astro.ex.ac.uk}  \\
School of Physics, University of Exeter, 
Stocker Road, Exeter, EX4 4QL \\}
\usepackage{amssymb}
\usepackage{amsmath}
\usepackage{graphicx}
\usepackage{epsfig}

\begin{document}
\date{\today}

\pagerange{\pageref{firstpage}--\pageref{lastpage}} \pubyear{0000}

\maketitle

\label{firstpage}
\begin{abstract}
We investigate the formation of GMCs in spiral galaxies through both agglomeration of clouds in the spiral arms, and self gravity.
The simulations presented include two-fluid models, which contain both cold and warm gas, although there is no heating or cooling between them.
We find agglomeration is predominant when both the warm and cold components of the ISM are effectively stable to gravitational instabilities. In this case, the spacing (and consequently mass) of clouds and spurs along the spiral arms is determined by the orbits of the gas particles and correlates with their epicyclic radii (or equivalently spiral shock strength). Notably GMCs formed primarily by agglomeration tend to be unbound associations of many smaller clouds, which disperse upon leaving the spiral arms.
 These GMCs are likely to be more massive in galaxies with stronger spiral shocks or higher surface densities. GMCs formed by agglomeration are also found to exhibit both 
prograde and retrograde rotation, a consequence of the clumpiness of the gas.
At higher surface densities, self gravity becomes more important in arranging both the warm and cold gas into clouds and spurs, and determining the properties of the most 
massive GMCs. These massive GMCs can be distinguished by their higher angular momentum, exhibit prograde rotation and are more bound. For a 20 M$_{\odot}$ pc$^{-2}$ disc, the spacing between the GMCs fits both the agglomeration and self gravity scenarios, as the maximum unstable wavelength of gravitational perturbations in the warm gas is similar to the spacing found when GMCs form solely by agglomeration.
\end{abstract}


\section{Introduction}
The accumulation of the ISM into giant molecular clouds (GMCs) represents the earliest stage in star formation. The properties of GMCs, such as turbulence and magnetic field strength regulate how star formation evolves on local scales (e.g. \citealt{McKee1999,Pudritz2002,Larson2003}), and are intrinsically linked to many of the issues in star formation, such as the time for stellar collapse (e.g. \citep{Elmegreen2007}). Thus understanding how GMCs form, and how their characteristics depend on the dynamics of galaxies and nature of the interstellar medium (ISM), is essential for progress in star formation. 
The formation and evolution of GMCs is also interrelated to the global properties of the ISM.
For example, GMC formation by coalescence is much more likely in a cold, clumpy medium, compared to a warm diffuse environment \citep{Dobbs2006}, where instabilities in the ISM are required.
In turn stellar feedback controls the ejection of hot gas back into the ISM and generates turbulence (e.g. \citealt{MacLow2004,Joung2006,deAvillez2007}). 

There have been numerous suggestions of how GMCs form (as described in a recent review by \citealt{McKee2007} and references therein). However they predominantly fall into two categories: either GMCs form by the agglomeration of smaller clouds of gas (e.g. \citealt{Field1965,Taff1972,Scoville1979,Casoli1982}), or through instabilities in the ISM. The latter include 
gravitational \citep{Cowie1981,Elmegreen1983,Balbus1985,LaVigne2006}, magnetic (either Parker instabilities \citep{Mous1974,Blitz1980} or MRI \citep{Kim2003}) or thermal instabilities \citep{F1965,Koyama2000,Stiele2006}. Another recent suggestion is that 
GMCs form in colliding or turbulent flows \citep{Vaz1995,Ball1999,Vaz2006,Heitsch2006}. This however requires a mechanism to produce these flows, which is most likely gravity \citep{Elmegreen2003}, spiral shocks or supernovae \citep{Koyama2000,Bergin2004}.

Coalescence of smaller clouds was first instigated to explain GMC formation, but the timescales of $10^8$ Myr for formation \citep{Kwan1979} were presumed too long to account for the observed properties of GMCs \citep{Blitz1980}. Furthermore star formation may be expected to disrupt the constituent clouds before a more massive GMC is assembled \citep{McKee2007}. By including spiral density waves, the time for formation by coalescence is reduced to 10's of Myrs \citep{Casoli1982,Kwan1987,Roberts1987}.

In more recent reviews, gravitational instabilities generally appear the preferred mechanism for GMC formation (e.g. \citealt{Elmegreen1990,McKee2007}). Theoretical analysis \citep{Elmegreen1982} and numerical simulations \citep{Kim2001,KOS2002} suggest that the Parker instability alone produces insufficient density enhancements, and GMCs can form by gravitational instabilities in relatively short timescales \citep{Elmegreen1990}. Predictions from gravitational analysis have also been compared to the observed masses and sizes of GMCs. In spiral galaxies, GMCs appear to be regularly spaced along the spiral arms. This spacing, and the mass of GMCs  can be interpreted in terms of gravitational perturbations to the gas \citep{Cowie1981,Elmegreen1983,Balbus1985,LaVigne2006}.
For a sound speed of 7 km s$^{-1}$, the estimated mass of a GMC is $10^6-10^7$ M$_{\odot}$, whilst the characteristic spacing is expected to be around 3 times the width of the spiral arm, corresponding approximately with observations \citep{Elmegreen1995}.  Finally, GMC formation by gravitational instabilities has further been endorsed by recent numerical simulations \citep{Kim2002,Kim2006,Shetty2006} which produce spacings and masses in line with the theoretical predictions. 

The problem of molecular cloud formation has been reopened in the last decade, with computational resources now enabling numerical simulations on both local \citep{Kim2002,Glover2007a,Glover2007b,Vaz2006,Heitsch2006} and galactic \citep{Shetty2006,DBP2006,DB2008,Tasker2008} scales. These simulations show the formation of GMCs by self gravity \citep{Kim2002,Kim2006,Shetty2006,Glover2007a}, turbulent or colliding flows \citep{Vaz2006,Glover2007b,Heitsch2006,Henne2008,Heitsch2008}, combined Parker and thermal instabilities \citep{Kosinski2007}, and Kelvin Helmholtz instabilities \citep{Wada2004,Wada2008}. In previous results \citep{Dobbs2006,DBP2006}, we showed that molecular clouds, and inter-arm spurs, can form as a result of cold gas passing through a spiral shock. The formation of this structure occurs without self gravity, and is still present when magnetic fields are included, although magnetic pressure acts to smooth out clumps in the spiral shock \citep{Dobbs2008}. 
The formation of clouds in these calculations bears most resemblance to the collisional models. Clumps of gas are forced together by the spiral shock, and agglomerate into larger structures, which become most discernible as interarm spurs extending from the spiral arms. Generally in simulations of grand design spirals, the spiral pattern is assumed to be long-lived (including the present work), although this may not be the case \citep{Merrifield2006,Shetty2007}.

This paper concentrates on two possibilities, the formation of GMCs by agglomeration, and by gravitational instabilities. In particular, we discuss the relative contribution to GMC formation for different initial conditions. When the surface density is lower, GMCs form by the agglomeration of smaller gas clouds as their orbits converge in spiral shocks, similar to the model by \citet{Roberts1987}. In this scenario, the spacing and size of clumps depends on the time (or equivalently distance) gas spends in the spiral arm, and interactions between clumps during orbit crossings. We show also the formation of GMCs form by gravitational instabilities for a high surface density disc with solely warm gas, and show that for a high surface density disc containing cold gas, both agglomeration and gravitational instabilities contribute to GMC formation.
We assess the differences in structure of the disc, in particular the separation of spurs, by performing MHD calculations with and without self gravity, varying the strength of the potential, disc mass, temperature and magnetic field strength.

\section{Calculations}
We use the 3D SPMHD code, a version of the SPH code originally written by Benz \citep{Benz1990} and \citet{Batesph1995}. The code has been extended to include magnetic fields \citep{Price2005,Price2004}. For the simulations presented here, the magnetic field is represented by Euler potentials (see also Section 2.2). The code also uses a variable smoothing length, such that the density $\rho$ and smoothing length $h$ are solved iteratively according to \citet{PM2007}. 
Artificial viscosity is included to treat shocks, with the standard values $\alpha=1$ and $\beta=2$ \citep{Monaghan1997}.

Long range gravitational forces are calculated using a binary tree algorithm \citep{Benz1990}. Over short scales, the gravitational forces are softened by setting the softening length equal to the smoothing length, as described in \citet{PM2007}. This method conserves both momentum and energy. Tests of the formulation of gravitational forces, and the implementation of gravitational softening are included in \citet{PM2007}. The code and the implementation of Euler potentials are described in more detail in \citet{PB2007} and \citet{Dobbs2008}.

\subsection{Thermal distribution of the gas}
The calculations presented here comprise of single phase and `two-fluid'  models. The gas in the single phase calculations is either cold (100 K) or warm ($10^4$ K) whilst in  the two fluid model, half the particles are allocated a temperature of 100 K and half $10^4$ K. In all cases, the calculations are isothermal, thus although the two fluid model includes cold and warm gas, there is no phase transition between the two components. 

This two fluid model is clearly a very simplified approach with numerous limitations. In particular, this model assumes that cold gas is widespread in the ISM, and enters the spiral shock. Recent calculations indicate that gas tends to heat up after passing a spiral shock \citep{Kim2008}, in which case the gas entering the shock may be warm. However in \citet{Wada2008}, there appears to be a considerable degree of cold gas in the midplane of the disc (Fig.~2) whilst some observations and theoretical analysis suggests cold HI or H$_2$ may be entering spiral shocks \citep{Vogel1988,Pringle2001,Gibson2007}. In calculations which include thermodynamics of the ISM \citep{DGCK2008}, we find that although gas entering the shock is usually warm, it tends to cool very quickly and produces clumpy, cold medium where agglomeration can occur. Future simulations with full MHD will be able to address this issue further.

In the two fluid models, the warm and cold gas generally separate out during the simulations, as the spiral shock induces much higher densities in the cold gas. In the agglomeration scenario, these results are not dependent on using a two fluid medium, since the structure in the cold gas is essentially the same as when only cold gas is used. In the low surface density case, where there is little substructure in the warm gas, and the cold clumps are largely separate from the warm gas, the pressure from the warm phase confines clumps and spurs to higher densities. Hence these features are easier to distinguish in analysis of the two fluid models. It is also useful to compare the structure of the disc for a given surface density when the gas is cold, warm and a mixture of both, and the latter is the most realistic distribution for the ISM of the 3 we investigate.

\subsection{Initial conditions and details of simulations}
We model an isothermal galactic disc between radii of  5 and 10 kpc. The gas is initially distributed uniformly with $z<150$ pc for the single phase cold calculations, and $z<400$ pc for the two fluid and  warm calculations. With time, the scale height is typically 20-100 pc for the cold gas and 300 pc for the warm \citep{Dobbs2008}. We perform calculations with 3 different surface densities (see Table~1), and for the majority of calculations, we use 4 million particles. The corresponding mass of the discs are $10^9$, $2 \times 10^9$ and $4 \times 10^9$ M$_{\odot}$. This gives a mass resolution of 250, 500 and 1250 M$_{\odot}$ for the 4, 8 and 20  M$_{\odot}$ pc$^{-2}$ surface density calculations respectively. In the appendix, a resolution study is included, with simulations of 1 and 8 million particles. All particles have the same mass, hence the two fluid simulations contain the same mass of warm and cold gas. 

The velocities in the plane of the disc follow a rotation curve corresponding to the disc component of the potential
\begin{equation}
\psi_{disc}=\frac{1}{2} v_0^2 \: log \bigg(\frac{r^2}{R_c^2}+\frac{z^2}{z_q^2}+1\bigg)
\end{equation} 
where $R_c$=1 kpc, $v_0=220$ km s$^{-1}$, and $z_q$=0.7 kpc is a measure of the disc scale height.
This produces an essentially flat rotation curve for the radii over which the particles are distributed. We also impose a velocity dispersion by selecting velocity perturbations from a Gaussian of mean 6 km s$^{-1}$. The velocity dispersion constitutes the $z$ component of the velocities for the particles.

The magnetic field in these simulations is initially toroidal and of uniform strength. A toroidal field can be described in terms of Euler potentials by
\begin{eqnarray}
\alpha_E & = & -B_0 \theta \\
\beta_E & = & \frac12 r^2
\end{eqnarray}
where $r^2 = x^2 + y^2 + z^2$ and $\theta = \cos^{-1}(z/r)$. The relative strength of the magnetic field is given by the
mean plasma beta, shown in Table~1. This is defined as the ratio of gas to magnetic pressure
\begin{equation}
\beta= \frac{P}{B^2 /2 \mu_0}=\frac{2 \mu_0 \rho_0 c^2}{|B|^2},
\end{equation}
where $\rho_0$ is the average density of the disc. We stress that although the magnetic field is uniform initially, there are large variations locally in the field strength and $\beta$ as the simulation progresses.

During the simulation, the gas is subject to an external potential. The potential includes a galactic halo \citep{Caldwell1981}  and the logarithmic disc component given in Eqn~1. The potential also incorporates a 4 armed spiral perturbation \citep{Cox2002,DBP2006}, with a pattern speed of $2 \times 10^8$ rad yr$^{-1}$, and a pitch angle of $15^o$. Calculations are performed with different strength potentials, as shown in Table~1. The strength $F$ is determined by the maximum radial force of the perturbation compared to the underlying disc potential, at a radius of 7.5 kpc, i.e.
\begin{equation}
F=max \bigg(\frac{d\psi_{sp}}{dr} \bigg|_{r=7.5}\bigg/\frac{d\psi_{disc}}{dr} \bigg|_{r=7.5}\bigg) 
\end{equation}
where $\psi_{sp}$ is the potential for the spiral perturbation (see also \citet{Roberts1969} and \citet{Shetty2006}).
We run the simulations for at least 250 Myr, or until runaway collapse occurs. 
\begin{table*}
\centering
\begin{tabular}{c|c|c|c|c|c|c|c|c|c|c|c|c|c|c}
\multicolumn{4}{l}{Two phase calculations} & & & & & & & & \\
 \hline 
 Model & T (cold) & $\Sigma$ & $\beta$ & Self & F & Q$_{c}$ & Q$_{h}$ & Spacing & r$_E$ & $\lambda_{max_c}$ &  $\lambda_{max_h}$ & M$_{clump}$  \\
& (K) & (M$_{\odot}$pc$^{-2}$) & & gravity & (\%) & & & (kpc) & (kpc) &  (kpc) &  (kpc) &  ($10^6$ M$_{\odot}$)  \\
 \hline
A & 100 & 4 & 4 &  Yes & 2 & 0.5 & 5 & 0.84 $\pm$0.04 & 0.45 & 0.066 & 6.6 & 0.5 \\
B & 100 & 4 & 4 &  Yes  & 4 & 0.5 & 5 &  1.05 $\pm$0.06 & 0.53 & 0.066 & 6.6 & 1\\
C & 100 & 4 & 0.4 & Yes & 4 & 0.5 & 5 & 0.91 $\pm$0.04 & 0.53 & 0.066 & 6.6 & 0.2\\
D & 100 & 4 & 4 &  Yes & 8 & 0.5 & 5 & 1.52 $\pm$0.09 & 0.73 & 0.066 & 6.6 & 3 \\
E & 1000 & 4 & 4 &  Yes & 8 & 0.5 & 5 & 1.33 $\pm$0.09 & 0.73 & 0.21 & 6.6 & 0.7\\
F & 100 & 4 & 4 &  Yes & 16 & 0.5 & 5 & 1.66 $\pm$0.04 & 0.85 & 0.066 & 6.6 & 4 \\
G & 100 & 4 & 4 & No & 4 & - & - & 1.04 $\pm$0.04 & 0.53 & - & - & 0.4 \\
H & 100 & 8 & 4 & Yes  & 4 & 0.25 & 2.5 & 1.12 $\pm$0.07  & 0.53 & 0.033 & 3.3 & 3\\
I  & 100 & 20 & 4 & Yes & 4 & 0.1 & 1 & 1.26 $\pm$0.06 & 0.53 & 0.013 & 1.3 & 20\\
\hline
\\[5pt]
\multicolumn{4}{l}{Single phase calculations} & & & & & & & & \\
 \hline 
 Model & T  & $\Sigma$ & $\beta$ & Self & F & Q$_{c}$ & Q$_{h}$ & Spacing & r$_E$ & $\lambda_{max_c}$ &  $\lambda_{max_h}$ & M$_{clump}$  \\
& (K) & (M$_{\odot}$pc$^{-2}$) & & gravity & (\%) & & & (kpc) & (kpc) &  (kpc) &  (kpc) &  ($10^6$ M$_{\odot}$)  \\
\hline
J & 100 & 4 & 1 & Yes  & 4 & 0.5 & - & 1.00 $\pm$0.03 & 0.53 & 0.066 & - & 1 \\
K & 10000 & 4 & 4 & Yes  & 4 & - & 5 & - & 0.53 & - & 6.6 & - \\
L & 100 & 20 & 4 & Yes  & 4 & 0.1 & - & - & 0.53 & 0.013 & - & - \\
M & 10000 & 20 & 4 & Yes  & 4 & - & 1 & 1.43 $\pm$0.06 & 0.53 & - & 1.3 & 8.5\\
N & 10000 & 32 & 4 & Yes  & 4 & - & 0.63 & 1.04 $\pm$0.04 & 0.53 & - & 0.8 & 7 \\
 \hline
\end{tabular}
\caption{Table showing the parameters used in these calculations.  All calculations use 4 million particles and in the two phase calculations, half the gas is distributed in the temperature indicated in the table, and half is $10^4$ K. $\beta$ is the ratio of thermal to magnetic pressure for the cold gas, and F is a measure of the strength of the spiral perturbation. r$_E$ is the epicyclic radius at $R=7.5$ kpc and $Q$ is the Toomre parameter. The calculation of these quantities is described in Sections~2.2 and 2.3. The spacing is calculated (also described in the text) after 265 Myr for all models except I (with time 130 Myr),  L (time = 180 Myr) and N (time = 85 Myr) where gravitational collapse prevents the simulations continuing. The final column is the mass of the most massive clump in a 5 kpc x 5 kpc section of the disc, using a clump-finding algorithm with a surface density threshold of 5$\Sigma$. In models K and L, regular large scale GMCs and spurs do not form within the time of the simulation.}
\end{table*}

\subsection{Calculation of $Q$, $\lambda_{max}$ and $r_E$}
This section describes the calculation of the parameters $Q$, $\lambda_{max}$ and $r_E$ listed in Table~1. These parameters are discussed in relation to the origins of GMCs in our simulations in Section~3.1.2. As mentioned in the introduction, the separations and masses of GMCs in galaxies can be compared to those predicted by theoretical models. We compute the expected separation and mass of large scale structures resulting from gravitational instabilities, as well as the stability of the disc, as measured by the Toomre parameter $Q$ \citep{Toomre1964}. However these are only approximations as we do not include a correction for the spiral shock \citep{Balbus1988}, and use the total column density, even in the two fluid models. We discuss any differences from these assumptions below. 

The $Q$ parameter is calculated by 
\begin{equation}
Q=\kappa c_s/\pi G \Sigma
\end{equation}
 where $\kappa=\sqrt 2 v_c/R$ ($v_c=220$ km s$^{-1}$ is the rotational velocity and the radius assumed is R=7.5 kpc), and $c_s^2=k_BT/\mu$ where $\mu=1.4$ is the mean molecular weight in these simulations. The disc is considered stable if Q $\gtrsim$ 1. For the two fluid models, we take the total surface density, but use the sound speed of the cold and warm gas to provide parameters $Q_c$ and $Q_h$ (Table~1). This is probably reasonable for the high surface density case where structures containing both cold and warm gas form along the spiral arms. However this treatment probably underestimates $Q$ in the low surface density cases (where the fluids are more separate) and the disc is more stable than suggested in Table~1. This calculation of $Q$ also neglects magnetic fields, which would modify $Q$ by a factor of $1+1/\beta$ (e.g. \citealt{Shu1992}).

For a thin disc, the dispersion relation arising from stability analysis of a differentially rotating disc is 
\begin{equation} 
\omega^2-\kappa^2+2\pi G \Sigma |k|-k^2 c_s^2=0
\end{equation}
(e.g. \citealt{Binney}). The separation of the clouds is expected to be approximately 
\begin{equation} 
\lambda_{max}=2c_s^2/G\Sigma,
\end{equation} 
the wavelength corresponding to the peak growth rate of perturbations \citep{Elmegreen1983,Balbus1985} whilst the minimum length at which perturbations become unstable for a thin disc, the Jeans length, is $c_s^2/G\Sigma$.
The mass enclosed within a cloud is 
\begin{equation}
M=\Sigma \bigg(\frac{\lambda_{max}}{2}\bigg)^2=\frac{c_s^4}{G^2 \Sigma}, 
\end{equation}
the Jeans mass for a thin rotating disc.
Again $\Sigma$ includes both the warm and cold gas in the two fluid models. However if only one component is included, this still produces a very different spacing from that of the low surface models. Also if we take the surface density of the spiral arms, as oppose to the average value,  $\lambda_{max}$ decreases still further for the cold gas. The warm gas does not experience such a strong increase in density in the spiral arms, but our $\lambda_{max}$ may be a high estimate. Overall though, $\lambda_{max}$ turns out to be very different from the spacing in calculations where gravitational instabilities are not believed to be dominating the large scale structure, but similar when gravitational instabilities are thought to be responsible. 

For calculations where the spacing is believe to correspond to formation by agglomeration, we compare the spacing to the epicyclic radius of the stellar orbits, a measure of the strength of the spiral potential.
The epicyclic radius is calculated for a 2D test particle simulation subject to the spiral potential alone, at a radius of  7.5 kpc. Particles are placed in a disc, and those with an average radius between 7.4 and 7.6 kpc over 2 rotational periods are selected. The epicyclic radius is then the average $(r_{max}-r_{min})/2$ of these particles (although this is a \textit{radial} measurement in the disc as oppose to along a spiral arm) .  

\section{Results}
We show the structure of the galactic disc for several simulations in Figs~1, 2 and 7.  Except for models K, N (no cold gas) and G (no self gravity), both agglomeration and accretion by gravity contribute to the formation of GMCs. When $\Sigma  \leq 8$ M$_{\odot}$ pc$^{-2}$, clouds and spurs form predominantly by agglomeration. However in models I, M and N, when $\Sigma= 20$ M$_{\odot}$ pc$^{-2}$, the gas becomes gravitationally unstable, and self gravity has a greater effect on the structure of the disc and the properties of GMCs. Thus we discuss the different surface density regimes separately in the next sections. 

\subsection{Formation of clouds by agglomeration (low surface density models)}
In Fig.~1, the galactic disc is displayed for models B, G, and D, which compare results with and without self gravity (a and b), and with a stronger potential (c). All three assume a two fluid medium. The last panel (d) shows the disc in model J, where there is only cold gas.
We see from b) and d) that the structure in the disc is dominated by the cold gas, rather than the warm, with essentially the same structure in the cold gas and two fluid simulations (J and B). 
The warm gas in these models does not experience significant density perturbations along the spiral arms.
Fig.~2 shows a disc with the same surface density, but all the gas is warm (run K). In this case, the disc is stable to gravitational instabilities, and the gas too warm for clump agglomeration to occur, so there is no substructure.

The top two panels of Fig.~1 compare simulations with (a) and without (b) self gravity. When there is no self gravity, spurs are still evident. With self gravity, the gas is arranged into more coherent clumps and spurs. However, the large scale structure, in particular the largest spurs which extend between the spiral arms, is very similar in both cases. Looking at a more detailed section of spiral arm (Fig.~3), there is even a 1 to 1 mapping between the spurs, with and without self gravity, although there is more fragmentation of the spurs without self gravity.

We interpret the formation of clumps and spurs in these simulations as due to agglomeration of gas in the spiral shock, rather than self gravity. In Fig.~4, we show the gas which constitutes one of the clumps in model B. The clump in the lower panel forms by the agglomeration of smaller clumps during the spiral shock passage. The same process occurs regardless of whether we include self gravity. 
However it is easier to distinguish spurs and spiral arms clumps, when self gravity is included.

By viewing a movie of gas flowing through a spiral arm, it is evident that clump agglomeration is not particularly efficient. Clumps do not always collide since the size scale of the clump is less than the width of a spiral arm, and collisions result in fragmentation as well as agglomeration. Self gravity increases the chances of agglomeration (see also \citealt{Kwan1987}) and leads to more distinct clumps (Figs~1, 3 and 4). 
It is apparent from Fig.~4 that the time to assemble a GMC structure from constituent clumps is around 50 Myr, i.e. comparable to the time for a spiral arm passage. 

We describe the process of agglomeration in more detail, as well as explaining why the spacing between clouds and spurs agrees with agglomeration as the main mechanism behind their formation in Section 3.1.2. Two further issues with the agglomeration model, the nature of the gas prior to the shock, and the dependence of cloud properties on numerical resolution are discussed in the appendices.
\begin{figure*}
\centerline{
\includegraphics[scale=0.4,bb=30 0 540 650,clip=true,angle=270]{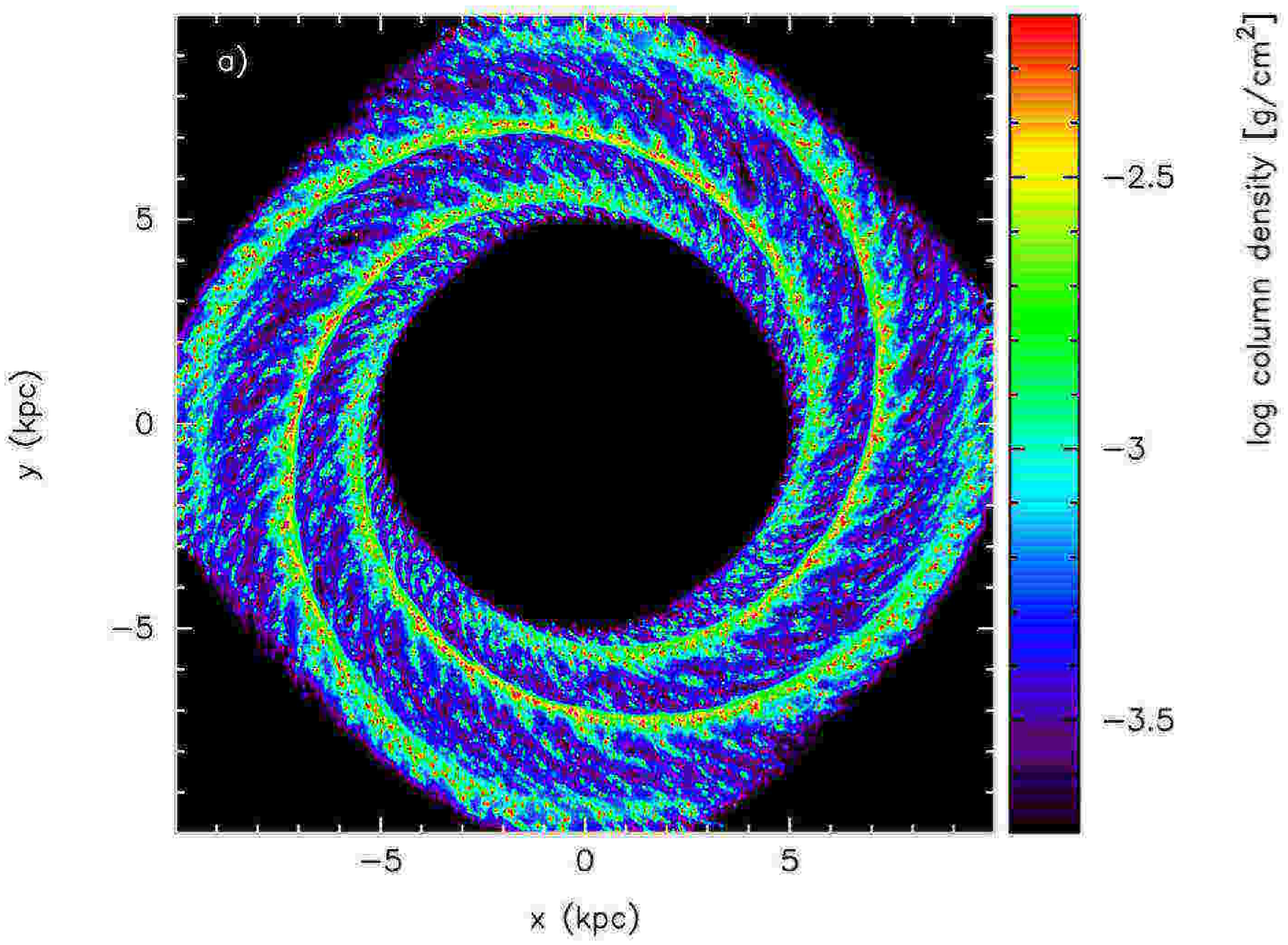} 
\includegraphics[scale=0.4,bb=30 0 540 650,clip=true,angle=270]{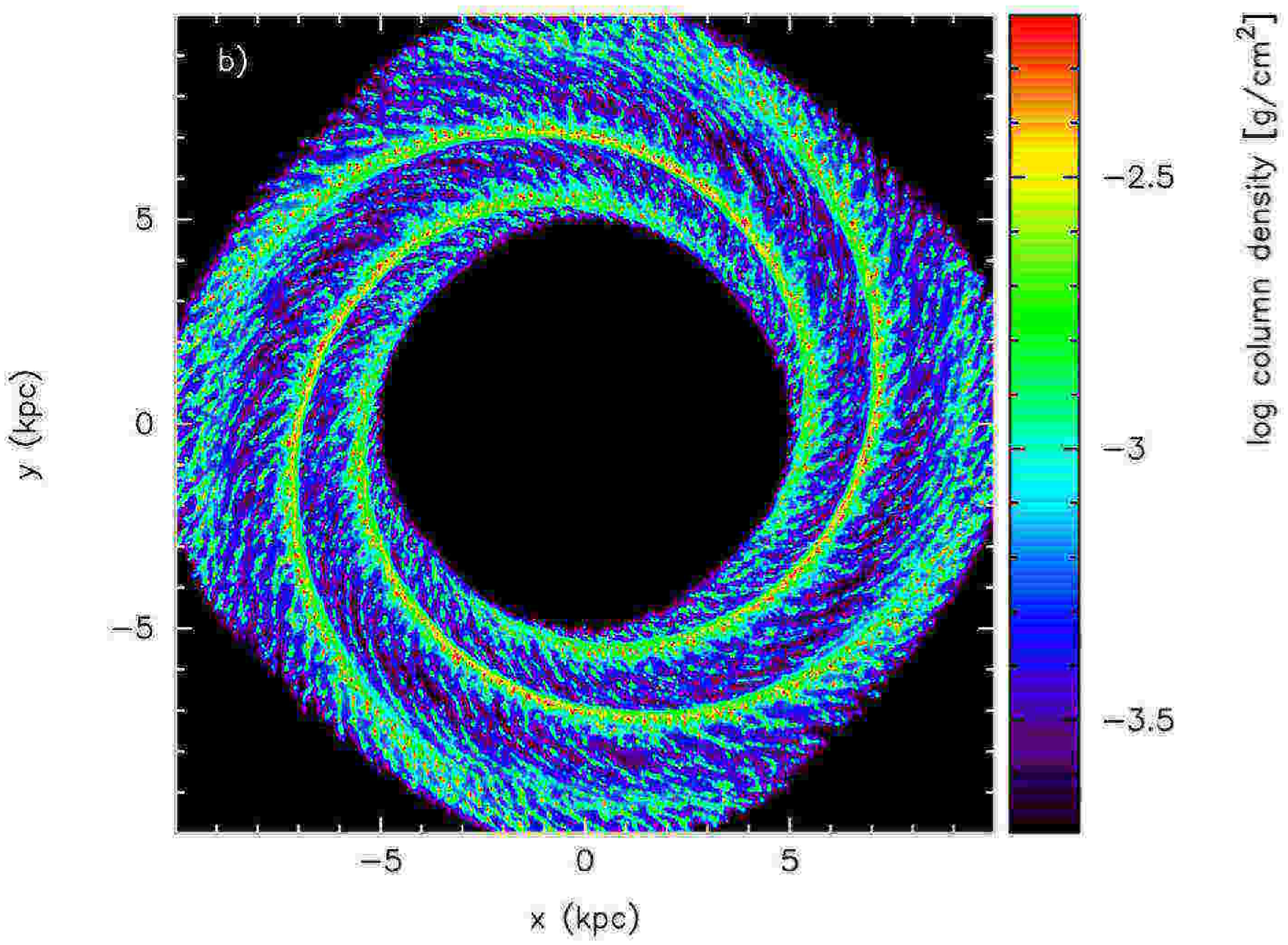}}
\centerline{
\includegraphics[scale=0.4,bb=30 0 540 650,clip=true,angle=270]{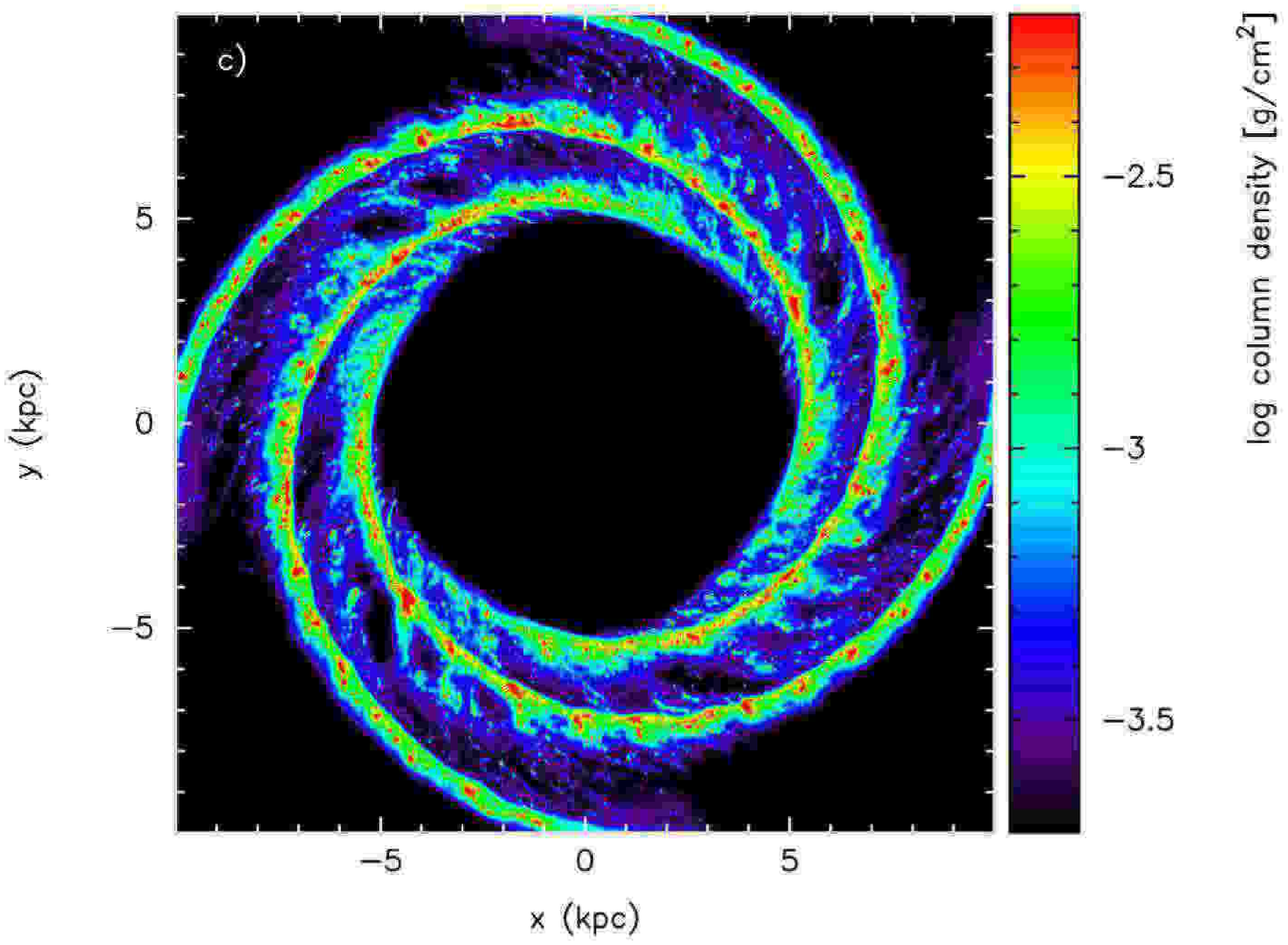}
\includegraphics[scale=0.4,bb=30 0 540 650,clip=true,angle=270]{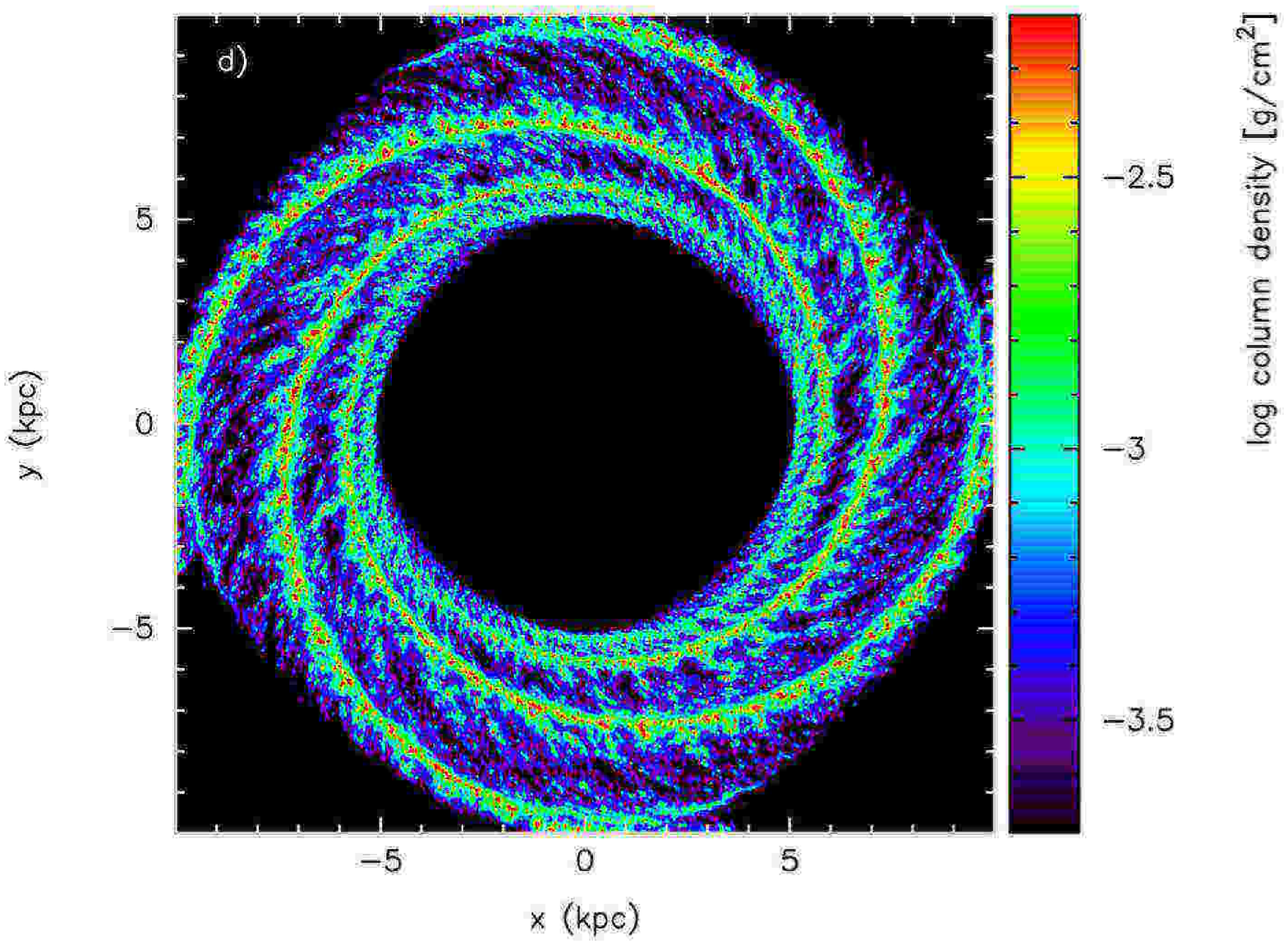}}
\caption{The galactic disc is shown for models B (a), G (b), D (c) and J (d) after 265 Myr. The top 2 panels compare results with (a) and without (b) self gravity. The spurs and structures in the arms are clearer when self gravity is included, although the overall structure is not significantly different. Both the lower panels (c and d) include self gravity. For (c), a stronger potential is used, and the spurs and clumps in the spiral arms are spaced further apart. The fourth panel (d) is from model J where all the gas is 100K. In all cases the disc surface density is 4 M$_{\odot}$ pc$^{-2}$.}
\end{figure*} 

\subsubsection{Wiggle instability}
An alternative explanation, which also does not require self gravity or magnetic fields, is that clumps and spurs are formed by the wiggle (Kelvin-Helmholtz) instability \citep{Wada2004,Wada2008}. Although this a possibility, we favour the agglomeration scenario as responsible for the large scale structure in our models. We still obtain structures at low resolution even though the Reynolds numbers are very low for resolving K-H instabilities \citep{Dobbs2006}. At lower resolution the instability appears weaker (fig. 4, \citet{Wada2004}) whereas spurs and clumps are actually more distinct in our simulations (see Appendix B). Secondly, the rotation curve of our disc is flat over the radii modeled, hence the disc should be relatively stable to K-H instabilities (although \citet{Wada2004}  point out that spiral streaming motions alone may induce instability). Thirdly agglomeration and merging of clumps is evident in simulations - even if KH instabilities are present initially, the gas evolves into clumps and the dynamics are then dominated by interactions between the clumps. Finally, as will be described in Section 3.1.2, we find a correlation between the spacing of clumps and the strength of the spiral potential, which is consistent with formation by agglomeration. It is not clear how the spacing varies if the structure arises from the wiggle instability.
  
 \subsubsection{Separation and masses of clouds}
In this section, we argue that the separation of clouds in the low surface density models is consistent with formation by agglomeration rather than gravitational instabilities.
Table~1 lists the separations between the main clouds/spurs along the spiral arms. These separations are calculated by selecting particles from a section of spiral arm. The particles are then transformed to co-ordinates parallel and perpendicular to the spiral shock, and particles binned into 5 pc sections over the coordinate parallel to the shock (thus the spacing includes the curvature of the spiral arm). We used two methods to obtain the spacing. Firstly, the average distance between the peaks along the spiral arm was calculated by selecting the main peaks by eye. This was repeated for the same region over each of the 4 spiral arms, and the average from the 4 spiral arms listed in Table~1.  The error is also calculated from the spread in these values.
We also applied Fourier analysis to the distribution of particles along the arm. For the models with stronger potentials (F=6\% and 12\%), the average spacing over the 4 arms is similar to those listed in Table~1, although with more spread. However for the lower strength potentials, the Fourier transform for a section of spiral arm sometimes produced peaks of similar magnitude. For these results it was inappropriate to extract a value for the spacing. Nevertheless we show in Fig.~5 some sample Fourier transforms where a reasonable signal was recovered for different strength potentials.  

For the low surface density models, the separations predicted from Jeans analysis of the warm gas are much too large compared to those in the simulations. The difference is not surprising since the warm component is gravitationally stable due to the low surface density. However, as evident in Table~1, the cold gas \textit{is} expected to collapse. Due to the dependence of the length scale in Eqn (8) on sound speed, the size scale of instabilities is only 10's of parsecs in the cold gas, and is therefore much smaller than the separation between the largest clumps.  Thus self gravity is important in the cold gas, but does not determine the large structure. The smaller spherical clumps shown in Fig~3, of size scales 50 pc or so, are comparable to the size scale associated with gravitational instabilities, and it is on these scales self gravity is having a more significant effect, and produces more coherent structures. The stability of the cold gas in the disc is discussed further in Section~3.1.3, and in particular why the cold gas is effectively stable and global collapse does not occur.

The spur separation is however found to increase with the strength of the spiral perturbation, or equivalently strength of the spiral shock (Table~1 and Fig.~5). This behaviour is contradictory with the Jeans analysis. If the spiral perturbation increases, we would expect $\lambda_{max}$ to stay the same or decrease, depending on whether we take the average surface density, or the surface density along the spiral arm. \citet{Shetty2006} find that the spacing resulting from gravitational instabilities is similar for different strength potentials. We also find that the spacing decreases for a calculation with 1000 K gas (E), corresponding to a decreasing shock strength, compared to that where the cold component is 100 K (D). If cloud formation is due to gravitational instabilities, the spacing would be expected to increase due to dependence on $c_s^2$ (Eqn.~8). Finally, when we double the disc mass, the spacing does not change significantly, as would be expected if the spacing is determined by the orbits. On the other hand if self gravity is dominant, we would expect the spacing to halve. 

Rather than self gravity determining the size and spacing of the spurs and clouds, we relate these properties to the convergence of orbits through the spiral shock. With gravitational instabilities, when a density perturbation occurs, there is a surrounding region (governed by the wavelength of the perturbation) of gas which will be subject to collapse. Analogous to this, in our models there is a locus of gas particles whose orbits will pass close to the initial perturbation. These gas particles are on slightly different orbits, so will have different velocities and angular momenta at a given point in the spiral arm. As they collide though, the gas particles (or small clumps of gas) exchange angular momentum\footnote[2]{Particles entering the shock are at the greatest extent of their orbits and have lower velocities compared with gas already in the shock. Thus these particles tend to gain angular momentum whilst material already in the shock loses angular momentum.} \citep{DBP2006}. The angular momenta and therefore velocities of the gas particles converge and  
locally a clump forms. Gas outside this locus, which has more disparate orbits, does not come sufficiently close to the density perturbation to experience much change in momentum. The locus of orbits over which gas can accumulate into a clump will depend primarily on the strength of the shock, i.e. the degree to which the orbits are perturbed by the shock.
\begin{figure}
\centerline{
\includegraphics[scale=0.35,angle=270]{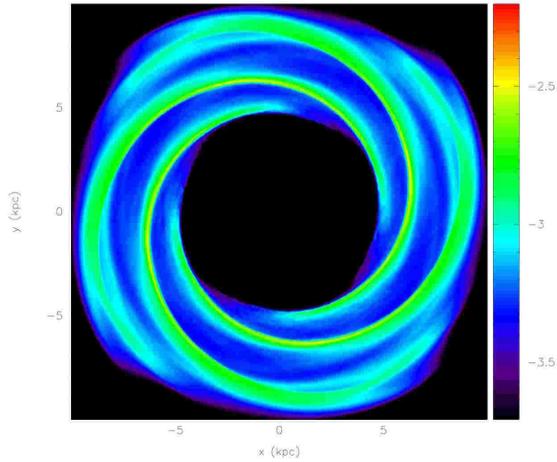}}
\caption{The column density of the disc is shown for model K after 235 Myr. All the gas is warm and the surface density is  4 M$_{\odot}$ pc$^{-2}$.  There is no substructure in the disc since the warm component is stable to gravitational instabilities, and the gas pressure prevents any clumpy structure evolving.}
\end{figure} 

\begin{figure}
\centerline{
\includegraphics[scale=0.47,bb=0 20 600 370,clip=true]{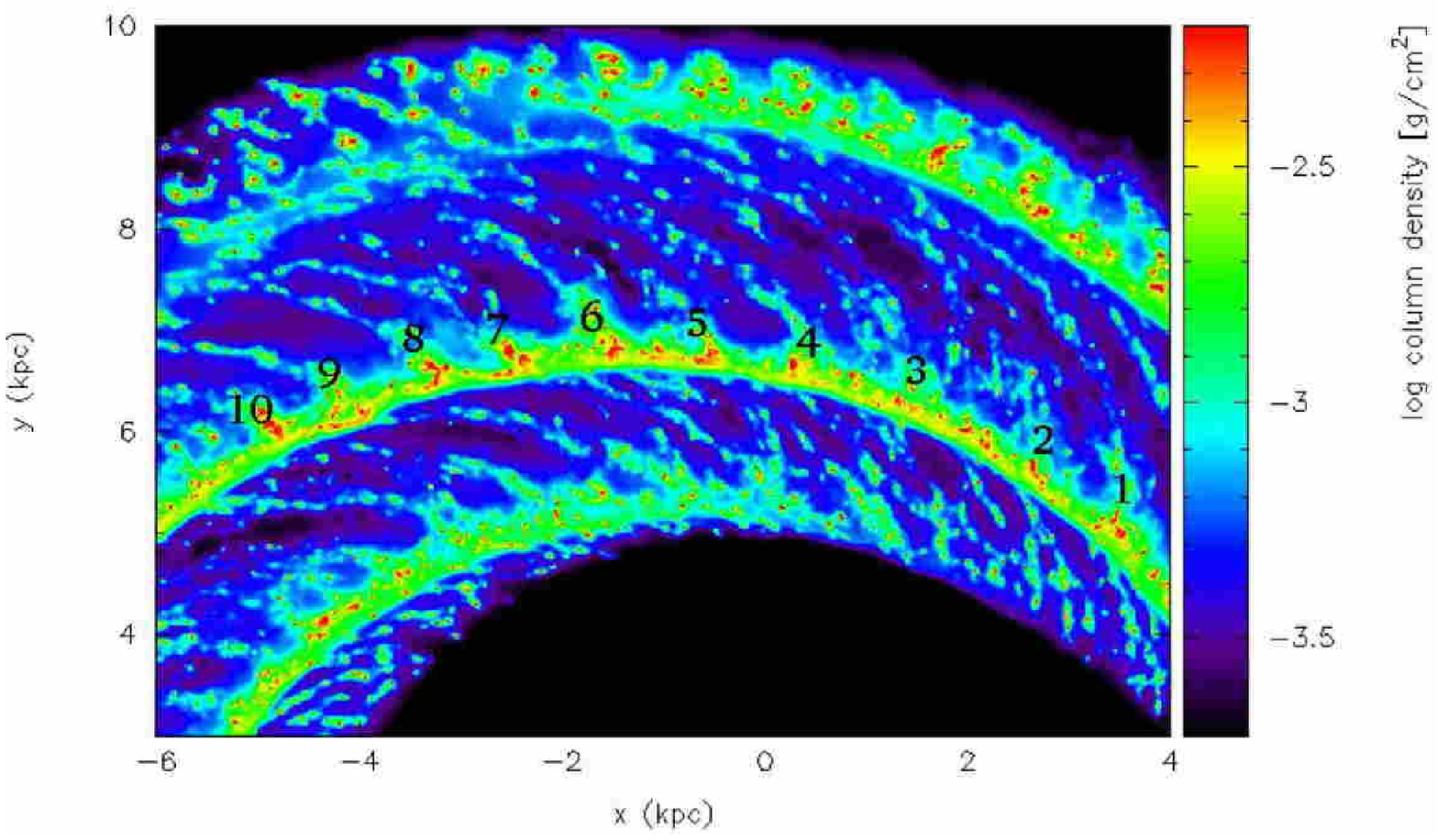}}
\centerline{
\includegraphics[scale=0.47,bb=0 20 600 370,clip=true]{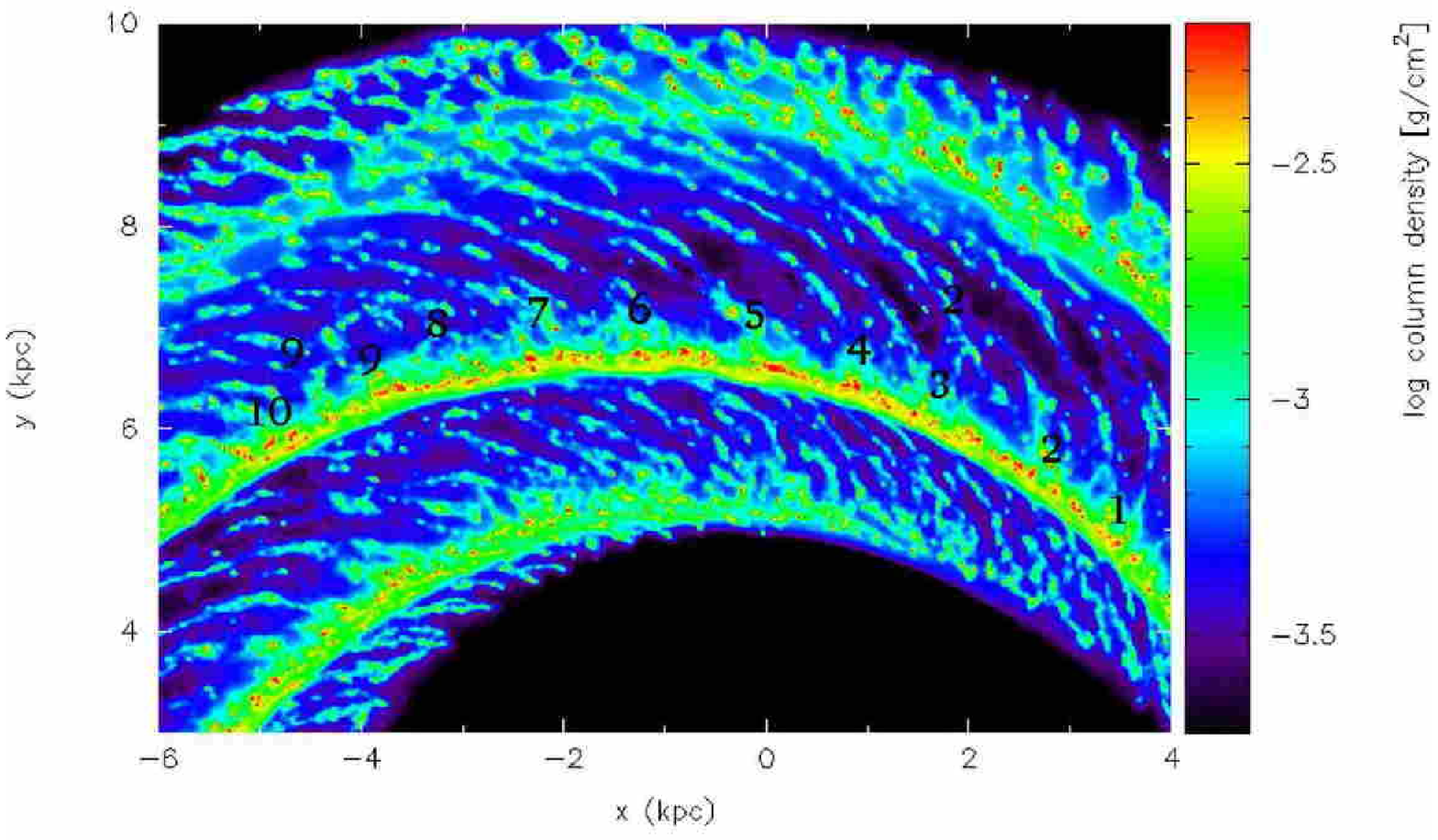}}  
\caption{A section of spiral arm is shown for models B and G, with (top) and without (lower) self gravity. The main features along the spiral are labelled. These are similar in each case, although with self gravity, the clumps and spurs are more coherent. Without self gravity these features are more dispersed and fragmented.}
\end{figure} 

\begin{figure}
\centerline{
\includegraphics[scale=0.22,angle=270,bb=50 50 550 660,clip=true]{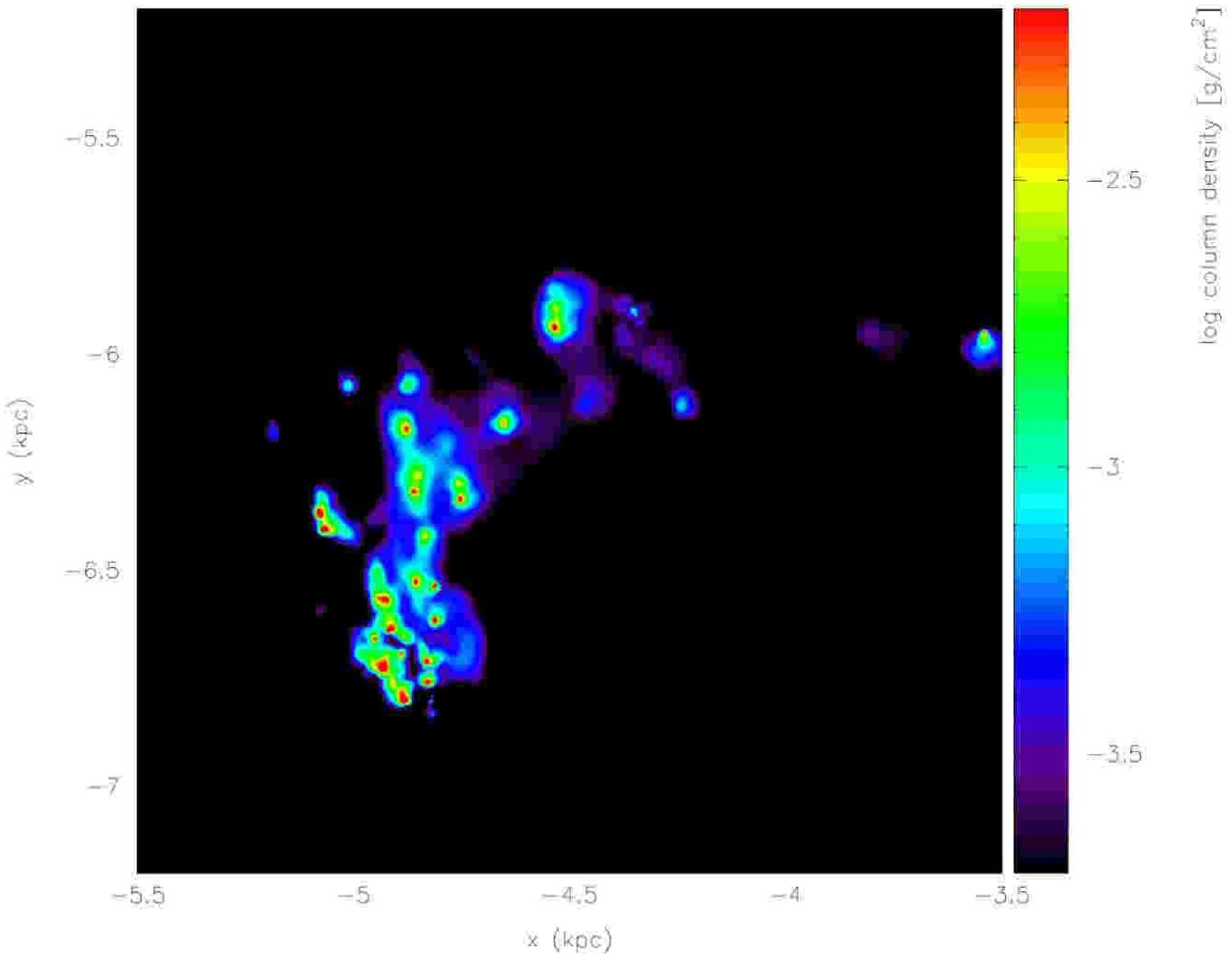}
\includegraphics[scale=0.22,angle=270,bb=50 50 550 660,clip=true]{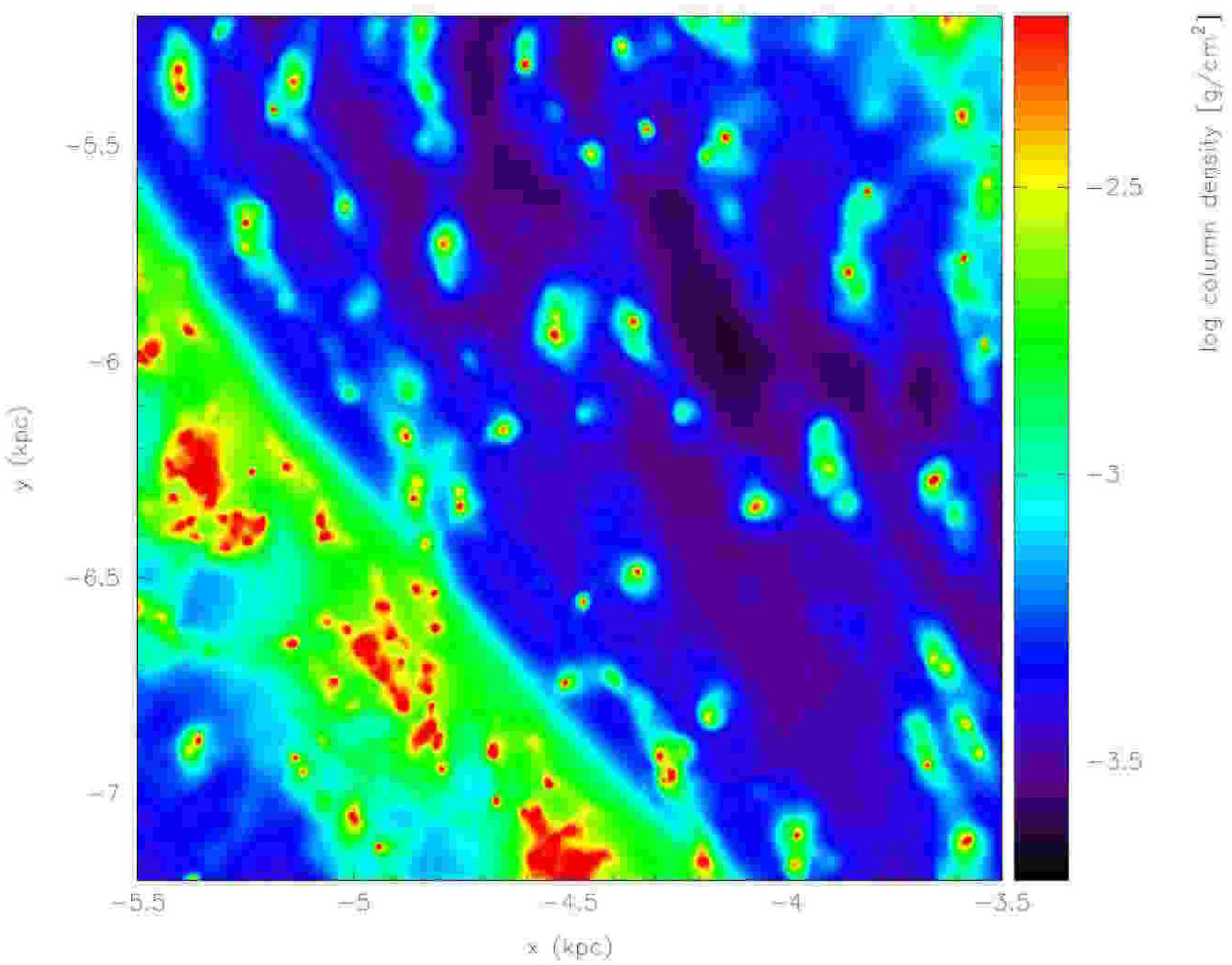}} 
\centerline{
\includegraphics[scale=0.22,angle=270,bb=50 50 550 660,clip=true]{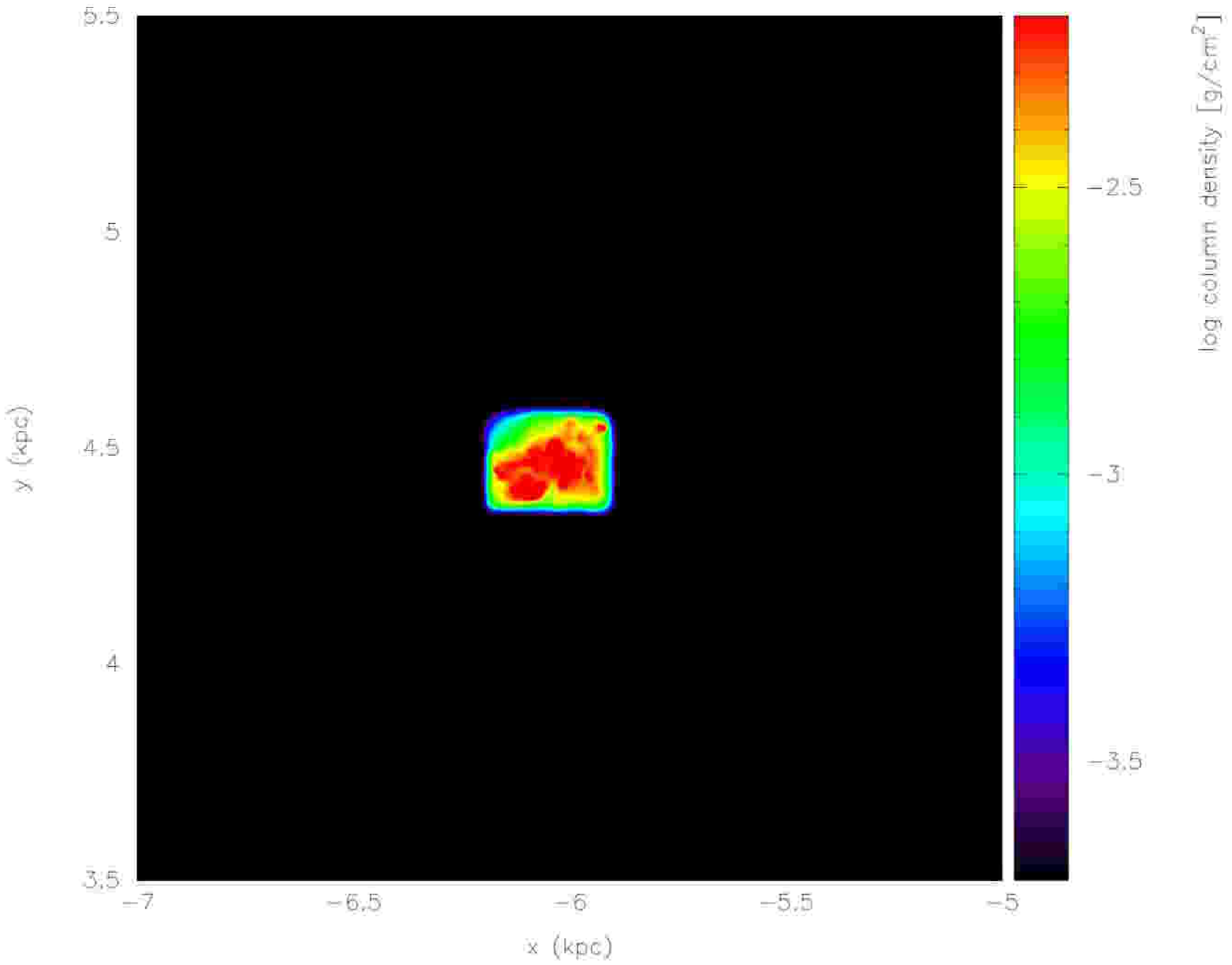}
\includegraphics[scale=0.22,angle=270,bb=50 50 550 660,clip=true]{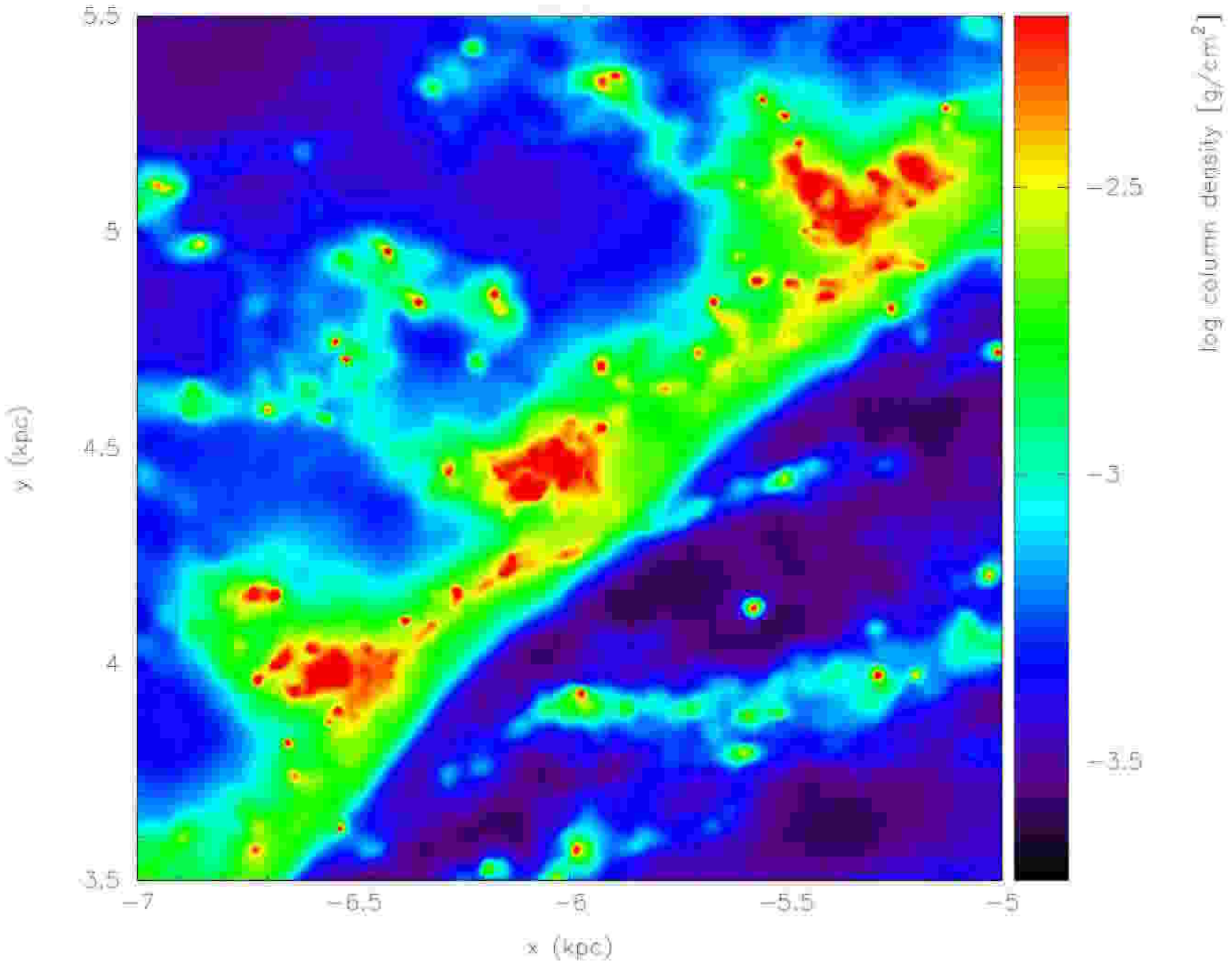}}
\caption{This figure shows the gas which constitutes a clump at an earlier time. The bottom panels show a clump from model B after 225 Myr, which contains a mass of $1.3 \times 10^6$ M$_{\odot}$. The gas which constitutes the clump is shown on the left and the surrounding material on the right. The top left panel shows the gas from the clump after 170 Myr (left), and again the surrounding region on the right. Thus the clump appears to form by the agglomeration of smaller clumps. This is in contrast to formation by gravitational instabilities in warm gas, where a perturbation develops from uniformly distributed gas.}
\end{figure}

\begin{figure}
\centerline{
\includegraphics[scale=0.4]{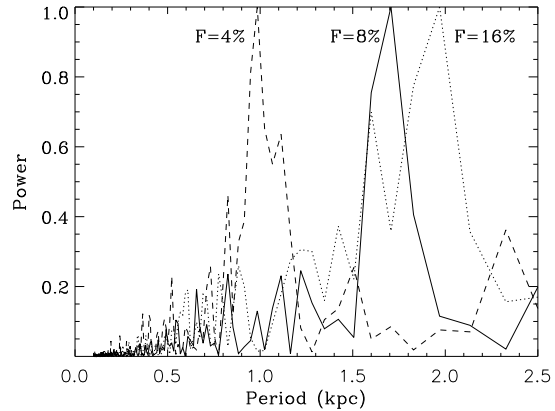}}
\caption{The Fourier transform is shown for the distribution of particles along a spiral arm for models B, D and F. In each case, the Fourier transform for the arm with the strongest signal is chosen (hence for the 8\% and 16\% cases the spacing is the largest of the 4 arms, since the efficiency of the shock in concentrating gas into large clumps will be highest to give the clearest signal). The period, i.e. distance between features, shifts to higher values as the strength of the shock increases.}
\end{figure} 
\begin{figure}
\centerline{
\includegraphics[scale=0.4]{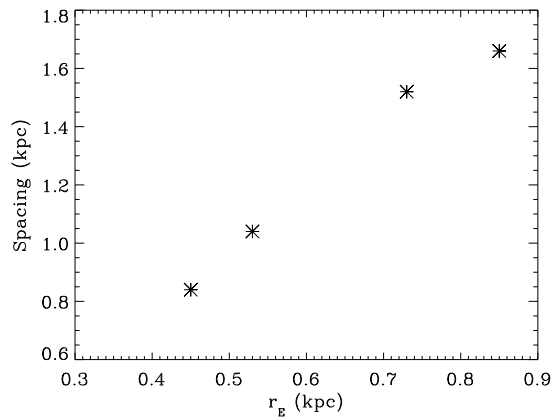}}
\caption{The average spacing between clouds (Table~1) is plotted against the epicyclic radius ($r_E$) associated with the potential. The spacing increases with the epicyclic radius (measuring the strength of the shock), and corresponds to approximately twice the epicyclic radius.}
\end{figure} 
One measure of the strength of the spiral shock is the epicyclic radius. This is related to the distance and time gas travels along the spiral shock (together with the pitch angle of the spiral potential and gas pressure), which in turn determines the size of the region from which the spiral shock can gather material into a single clump.
In Table~1, we show the epicyclic radius calculated at a radius of 7.5 kpc from the orbits of particles subject to the potential alone. Fig.~6 shows the spacing from Table~1 plotted against the epicyclic radius. 
We see that the spur spacing tends to be about twice the epicyclic radius (also equivalent to the maximum radial extent of the gas as it passes through its orbit). For the stronger shock, the orbits are more perturbed, so the epicyclic radius and spacing between the clumps becomes larger. However with greater thermal and/or magnetic pressure (models E and C), the spiral shock becomes weaker. With a weaker shock, there is less orbit crowding, and fewer interactions between gas particles or clumps. Interactions also produce a smaller transfer of angular momentum, since the difference in velocities of gas within the spiral shock, and compared to gas entering the spiral shock, is less. Hence there is a smaller locus over which particles are subject to a sufficient change in angular momentum to be concentrated into a clump, and the spacing between the clumps is smaller.

Table~1 also includes the mass of the largest clump along a section of spiral arm for each model (determined using a clumpfinding algorithm, see Section~3.3). The typical cloud masses are $\sim 10^6$ M$_{\odot}$, which is much larger than the Jeans mass for the cold gas ($\sim$ 5000 M$_{\odot}$). For cloud formation by gravitational instabilities, the Jeans analysis predicts that discs with larger surface densities should contain lower mass clouds, whilst a stronger shock will produce lower or the same mass clouds depending on whether the average or spiral arm surface density is considered. For agglomeration, both an increased surface density or a stronger spiral potential would be expected to lead to higher mass clouds, as found for these results (runs D, F and H).

\subsubsection{Gravitational stability of the cold gas}
As mentioned in the previous section, $Q<1$ for the cold gas, hence the cold component is expected to be globally unstable to gravitational instabilities. Previous simulations indicate that an unmagnetized disc can be unstable even if Q=2 \citep{Shetty2006}. Although self gravity influences the density and morphology of clumps in the disc, runaway collapse does not occur in the low density models. Collapse may be prevented in our models by supersonic motions, magnetic fields or insufficient numerical resolution. However we stress again that the size scale of gravitational instabilities in the cold gas in our simulations is of order 10Õs parsecs, and thus much less than the scale of the features found in warm gas in \citet{Shetty2006}. Therefore, ignoring stellar feedback, self gravity is not likely to significantly change the large scale distribution of clouds and spurs in our models even if collapse occurs.

To investigate the disc stability explicitly, we performed further simulations without the spiral perturbation. Without magnetic fields, the cold component of a two fluid disc with $\Sigma=4$ M$_{\odot}$ pc$^{-2}$ is unstable. As would be expected, the initial velocity dispersion is not sustained on small scales and collapse occurs. However with a magnetic field of the same strength as model B, the disc is stable. The overall outcome, i.e. that runaway collapse does not occur when magnetic fields are present, is therefore the same whether or not the spiral potential is applied.

When magnetic fields are included, runaway collapse is prevented by the overall pressure of the gas, which comprises both a thermal and magnetic component. Although $\beta>1$ globally, locally the magnetic pressure can be much higher than the thermal pressure, and since the magnetic fields are fairly disordered in the clumps, the magnetic pressure is relatively isotropic. We previously found $B \propto \rho^{0.7}$ \citep{Dobbs2008}, so the magnetic field is strong ($\sim$ 100 $\mu$G) in dense regions. Furthermore $\beta$ is then $<1$, typically between 0.001 and 0.1. To illustrate the role of magnetic pressure, we consider the Jeans length of a local spherical clump. For dense gas in our simulation, with $\rho=10^{-19}$ g cm$^{-3}$, the Jeans length $\lambda_J=(c_s^2 \pi/G \rho)^{1/2}$ is $ \sim$ 0.5 pc in the cold gas. For the magnetic Jeans length however, the sound speed is replaced by the Alfv\'{e}n speed, i.e. $\lambda_B=(v_a^2 \pi/G \rho)^{1/2}$ where $v_A$=$\sqrt{2} c_s/\sqrt{\beta}$ \citep{Chand1961}. This dependence on $\beta$ increases the local Jeans length of most gas at this density to around 10-100 pc. Thus collapse does not continue below these scales, and is prevented by magnetic and thermal pressure. However for the 20 M$_{\sun}$ pc$^{-2}$ calculation (model I), the densities become sufficiently large to overcome the pressure, hence runaway collapse halts the calculations. We could use a similar argument by taking the local value of $Q$, again modified to include magnetic fields. \citet{Shetty2006} similarly find that collapse occurs earlier (within an orbit) without magnetic fields, but this is rather a consequence of a strong ordered field which prevents collapse along the magnetic field lines.

Nevertheless, even in the lower surface density models there is a distribution of $\beta$ for a given density. At higher resolution, there may be clumps with $\beta \lesssim 1$ which could collapse, but are not resolved in our simulations.  We performed simulations without magnetic fields at different resolutions. The peak density in the disc increases with resolution, and consequently at lower resolution simulations are able to run for slightly longer before collapse occurs. Thus the combined thermal and magnetic pressures primarily prevent collapse, but numerical resolution is also important.

For the low surface density models, self gravity has only a small effect on the gas before it enters a spiral shock (Appendix A). Although the density increases in the shock, this is counteracted by an increase in the velocity dispersion of the gas due to collisions between clumps \citep{Bonnell2006,DB2007}. In the most dense clumps though, self gravity becomes more important and gas reaches even higher densities. The dense clumps are those which are most bound, particularly as they move away from the spiral shock and the velocity dispersion decays. Clumps which become gravitationally bound during or after the spiral shock are then supported by magnetic and thermal pressure, and as mentioned above have field strengths in excess of 100 $\mu$G. However, as mentioned previously, gas in the bound clumps may undergo runaway collapse at higher resolution. Furthermore, we also neglect thermodynamic processes and non-ideal MHD effects such as ambipolar diffusion, which are likely to promote collapse.

As we show in Section~3.3.2, the largest GMCs which form in these low surface density calculations tend to be unbound associations of smaller clumps. The overall picture we present is that self gravity is likely to be important with regards to small, dense clumps within a GMC, but it is the action of the spiral shock which focuses the clumps into a GMC (GMA (giant molecular association) may well be more appropriate). Although the most dense clumps within the GMC may become bound, supported by magnetic fields and supersonic velocities, the whole GMC (or GMA) is not necessarily bound. 
  
\subsection{GMC formation by agglomeration and gravitational instabilities (high surface density models)}
We now consider discs with a higher surface density, where self gravity is likely to have a more important role in the formation of GMCs and their resulting properties. Fig.~7 shows the disc structure with a much higher surface density, of 20 M$_{\odot}$ pc$^{-2}$. The top panel shows a single phase simulation containing only $10^4$ K gas (model M), after 265 Myr. The clumps and spurs are purely due to self gravity, as simulations of warm gas do not show such features in the absence of self gravity \citep{Dobbs2006}. The spacing is roughly in agreement with the predictions of \citet{Elmegreen1983}, and not too dissimilar to \citet{Shetty2006}. As a further test, a simulation with twice the disc mass was performed (model N), and the spacing between clumps became clearly smaller, as would be expected from Eqn~8. Thus overall these simulations with warm gas agree with those of \citet{Shetty2006}, in finding that GMCs form by self gravity.

For the 20 M$_{\odot}$ pc$^{-2}$ density calculations which include cold gas, gravitational collapse halts the calculations after 130 Myr (model I) and 180 Myr (model L).  
The middle panel of Fig.~7 shows a calculation with cold and warm gas after 130 Myr (model I). The structure of the disc is different to single phase calculations with cold or warm gas after this stage. Unlike the lower surface density calculations, the warm gas experiences density perturbations along the spiral arms. Large coherent complexes are located along the spiral arms, which consist of envelopes of warm gas, and dense cores of cold gas. These large complexes have masses of $10^7$~M$_{\odot}$ or more.

The massive GMCs in the two fluid calculation, model I, form by a combination of self gravity, and agglomeration between clumps of cold gas. The structure of the disc for model I deviates from the low surface models B and G, with the presence of larger clouds along the spiral arm, whilst the spiral shock is less evident. The spacing between the most massive complexes is however compatible with either gravitational instabilities in the warm gas or the agglomeration scenario. The complexes are separated by $\sim 1.2$ kpc, which is similar to both $\lambda_{max}$ for the warm gas, and the separation in the low surface density models, B, G and H. There is similarity in the structure for models I and L, so it is plausible that gravitational perturbations in the warm gas are seeded by the most dense cold clumps. At later stages of the simulation, cold clumps tend to be confined to the minima of the gravitational potential due to the warm gas, increasing in mass by the agglomeration and accretion of smaller clumps of cold gas. As described in the next section, the properties of the massive clouds in this simulation suggest that self gravity has a much more dominant role in determining their structure than for the low surface density calculations. 

With only cold gas (model L, Fig.~7 lower panel), there is more substructure in the gas, and gravitational collapse occurs on very small scales since $\lambda_{max} \sim$ 10-20 pc.  However, self gravity also has a much greater role in organising the large scale structure than for the low surface density models, as can be seen by comparing Fig.~7 (lower) and Fig.~1d), which also has only cold gas, but a surface density of  4 M$_{\odot}$ pc$^{-2}$. Thus large scale structure is present along the spiral arms, similar to the two fluid model (I), and again self gravity and agglomeration contribute to the formation of GMCs. However for the two fluid model (I), the warm gas appears to help concentrate the cold gas into singular large clumps, which smoothes out some of the structure in the disc. 

\begin{figure}
\centerline{
\includegraphics[scale=0.32,angle=270]{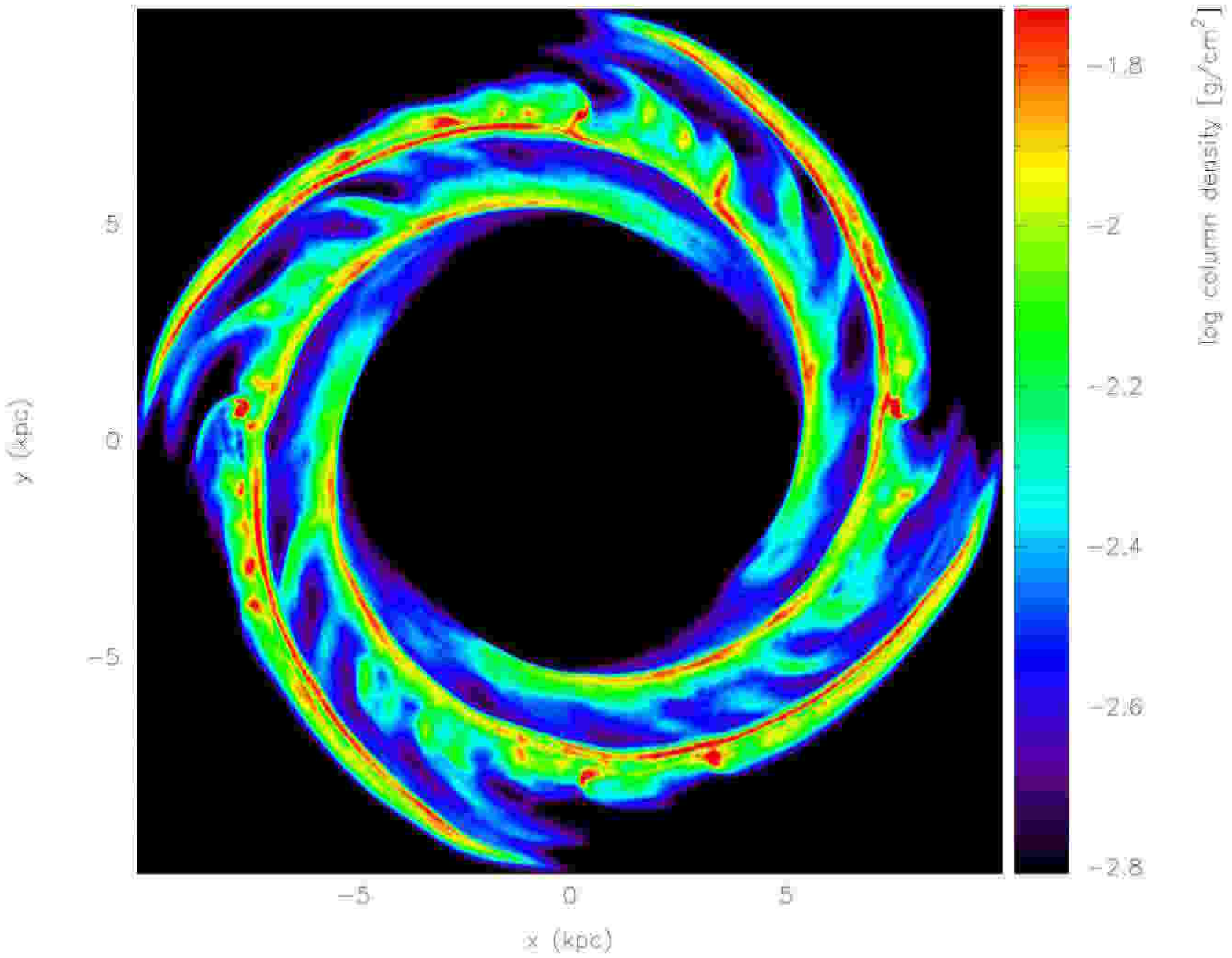}} 
\centerline{
\includegraphics[scale=0.32,angle=270]{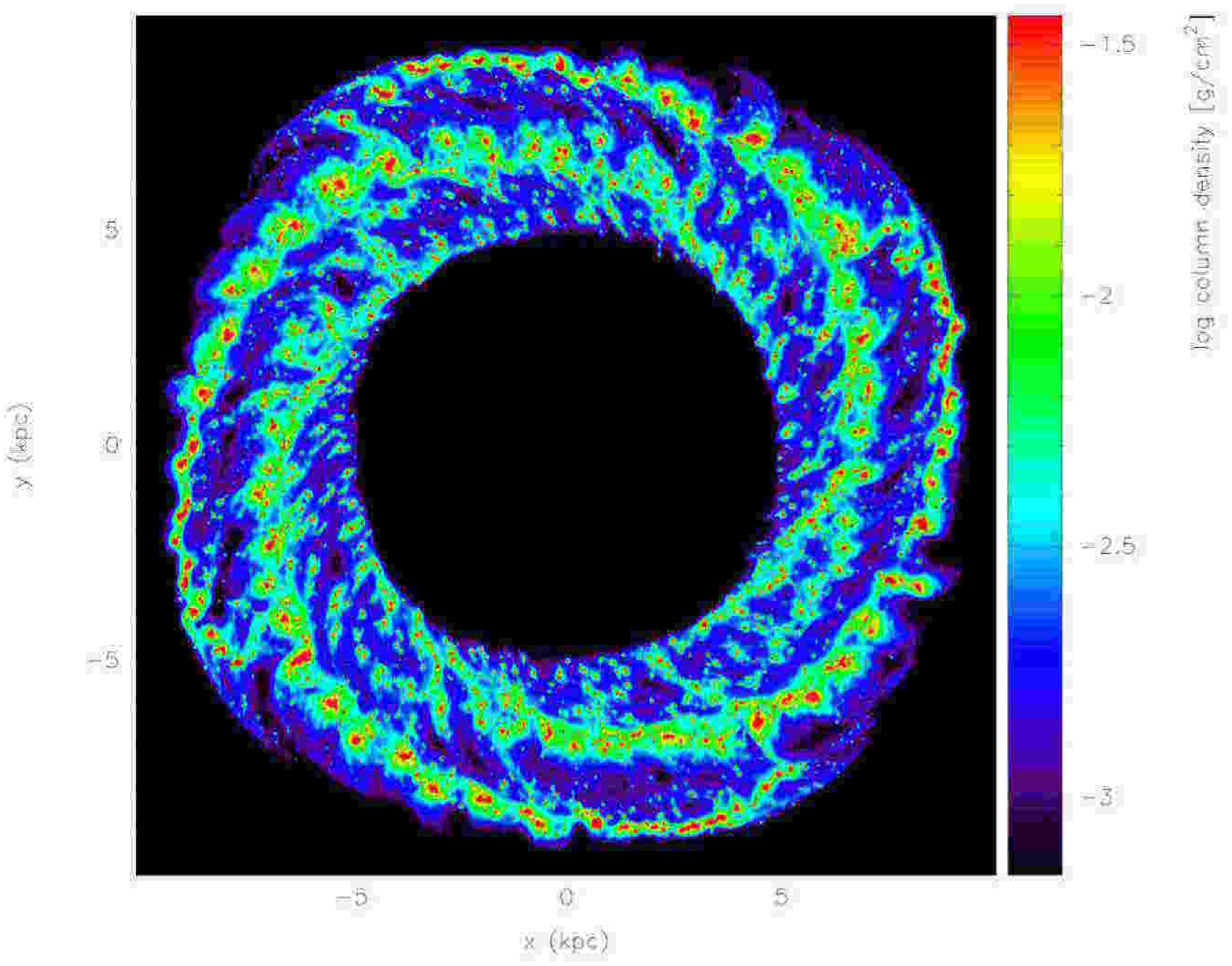}} 
\centerline{
\includegraphics[scale=0.32,angle=270]{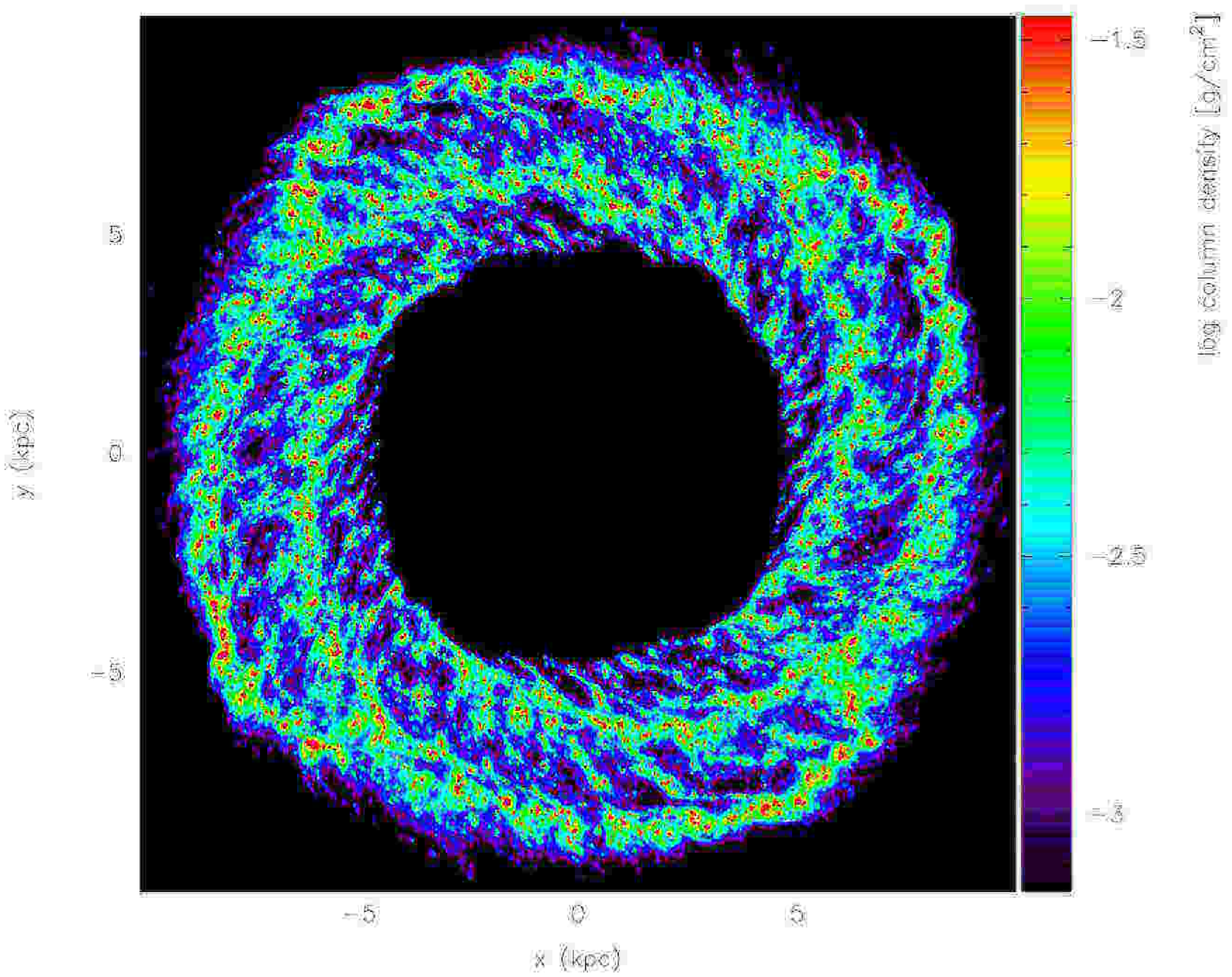}} 
\caption{The column density of the disc is shown for models M (top), I (middle) and L (lower) after 265, 130 and 130 Myr respectively. The gas is either all warm (top panel), half warm and half cold (middle panel) or all cold (lower panel). In these simulations, the disc mass is 5 times larger than in Fig.~1. Consequently the warm gas is unstable to gravitational instabilities. With no cold gas, the substructure in the top panel is solely due to self gravity. In the middle panel, both warm and cold gas are included, and the large complexes are presumed to form through gravitational instabilities in the warm gas, and agglomeration between cold clumps. In the lower panel, all the gas is cold, and although the disc appears similar to the middle panel, the coherent complexes present in model I along the spiral arms are largely absent, and there is more substructure.}
\end{figure}

\subsection{Properties of clumps formed in these simulations} 
In this section we compare the properties of the clumps formed in the low and high surface density cases. These could potentially be compared to observations and give an indication of the main processes contributing to GMC formation in a particular galaxy. In particular, we consider the angular momentum of clouds in our models and compare with observations of M33.

To select clouds, we applied a clump-finding algorithm to these simulations. The algorithm selects regions with surface densities of $5\Sigma$, where $\Sigma$ is the initial average surface density, with a resolution of 10 pc. Only clumps with $\geq 30$ particles are retained (equivalent to 7500, 15000 and 37500~M$_{\odot}$ for the 4, 8 and 20  M$_{\odot}$ pc$^{-2}$ surface density calculations respectively). Although we do not include molecular hydrogen formation in these calculations, the densest regions are likely to correspond to GMCs.

\subsubsection{Angular momentum of GMCs and implications for the ISM}
\citet{Rosolowsky2003} have determined the angular momentum from clouds in M33 and compared these values with those expected from several theories of GMC formation. Their values of angular momentum are lower than expected if GMCs form by gravitational instabilities. Furthermore 40\% of their clouds show retrograde motion, which cannot be explained if self gravity is predominant. They suggest that magnetic braking may play a role in removing angular momentum from GMCs as they form. 

Fig.~8 shows the specific angular momentum of the clouds found in models B and I (with respect to the cloud's centre of mass). We see that in both cases the GMCs exhibit prograde and retrograde rotation. The distribution of angular momentum and ratio of prograde and retrograde clouds is similar in models B and G, i.e. with and without self gravity, with approximately twice as many prograde clouds as retrograde. Retrograde clouds arise through collisions between clumps (see also \citealt{Tasker2008}), and the frequency of retrograde clouds which arises in these models is an indication of the clumpy nature of the gas in the simulations. Retrograde motions are not expected if the gas is treated solely as a fluid, and accordingly  all the GMCs (formed by gravitational instabilities) in model M (with only warm gas) were found to exhibit prograde rotation, although there are far fewer objects. Thus observations of retrograde clouds in M33 suggest that at least some component of the ISM is clumpy.
The magnitude of the specific angular momentum in models B and G, where clumps form mainly by agglomeration, does not exceed 100 km s$^{-1}$ pc and is comparable to the values obtained by \citet{Rosolowsky2003} for M33.

When the surface density is 20 M$_{\odot}$ pc$^{-2}$, the distribution of clouds of mass $\lesssim 4 \times 10^6$  M$_{\odot}$ is similar, and again there are twice as many prograde clouds as retrograde. However there is a surplus of prograde clouds with a high angular momentum. These are the most massive clouds where self gravity is significantly contributing to the growth of the cloud. The rotation of the clouds is visibly evident from the presence of spiral tails of gas (Figs.~7 and 10). Again these most massive clouds are associated with instabilities in the warm component of the ISM, which behaves as as fluid. 
 
\subsubsection{Mass spectrum and gravitational stability of the GMCs}
We also compared the boundedness of clouds in these calculations. As expected, in the simulations with a higher surface density, the clumps are more bound. For model B, where $\Sigma=4$ M$_{\odot}$ pc$^{-2}$, the virial ratio ($\alpha$) is between 1 and 6 for the clumps of mass $>10^4$ M$_{\odot}$, so these clouds are unbound ($\alpha=5 \sigma^2 R/ 3GM$ where $\sigma$ is the 3D velocity dispersion, $R$ is the radius of the clump and $M$ the mass).  Any marginally bound clumps in model B are sufficiently supported by magnetic and thermal pressure for the calculation to continue. With model I, where $\Sigma=20$ M$_{\odot}$ pc$^{-2}$, $\alpha$ lies between 0.7 and 2 for clouds of mass $>10^5$ M$_{\odot}$. In model I, the magnetic and thermal pressure are not sufficient to prevent collapse in the most bound clouds, and the calculation stops. In the low surface density runs, the clouds are more bound when self gravity is included. Without self gravity (model G), $\alpha$ is distributed about a mean of approximately 5 for clumps $>10^4$ M$_{\odot}$. 

The distribution of clouds also changes with self gravity, as the simulations with self gravity contain more interarm clouds. Clouds formed in the spiral arms are more bound, and remain intact for longer in the interarm passages. We also found that the clouds in model I, with the higher surface density, produced a shallower mass spectrum (Fig.~9), with $dN/dM \propto M^{-1.75}$ compared to $M^{-2}$ for model B (self gravity and low surface density) and $M^{-2.1}$ for model G (no self gravity). At high masses, self gravity increases the agglomeration of clumps and accretion onto clouds, so there are more massive clouds and a shallower slope. A similar result was produced recently in simulations of prestellar cores \citep{Dib2008}. 

Fig~10 shows individual clouds from the low and high surface density calculations, B and I. As mentioned in Section~3.1.2, clouds formed by agglomeration also tend to consist of numerous clumps dispersed throughout the GMC. By contrast, in the high surface density run where the GMCs are more bound, the GMCs are much more centrally concentrated. 

Generally, the properties of clouds formed in model I, where self gravity plays a strong role in the formation of GMCs are most similar to those observed for the Milky Way (i.e. lower $\alpha$ and the shallower mass spectrum). Even the velocity sizescale relation of the clouds resembles the $\sigma \propto r^{0.5}$ observed law, whilst there is no clear relation in the other simulations. However, the properties of clouds in model I are dissimilar to M33, where there are no $10^6$ M$_{\odot}$ GMCs. The mass spectrum is also steeper \citep{Blitz2007}, and the GMCs all have relatively low specific angular momenta.
\begin{figure}
\centerline{
\includegraphics[scale=0.55,bb=50 300 500 600,clip=true]{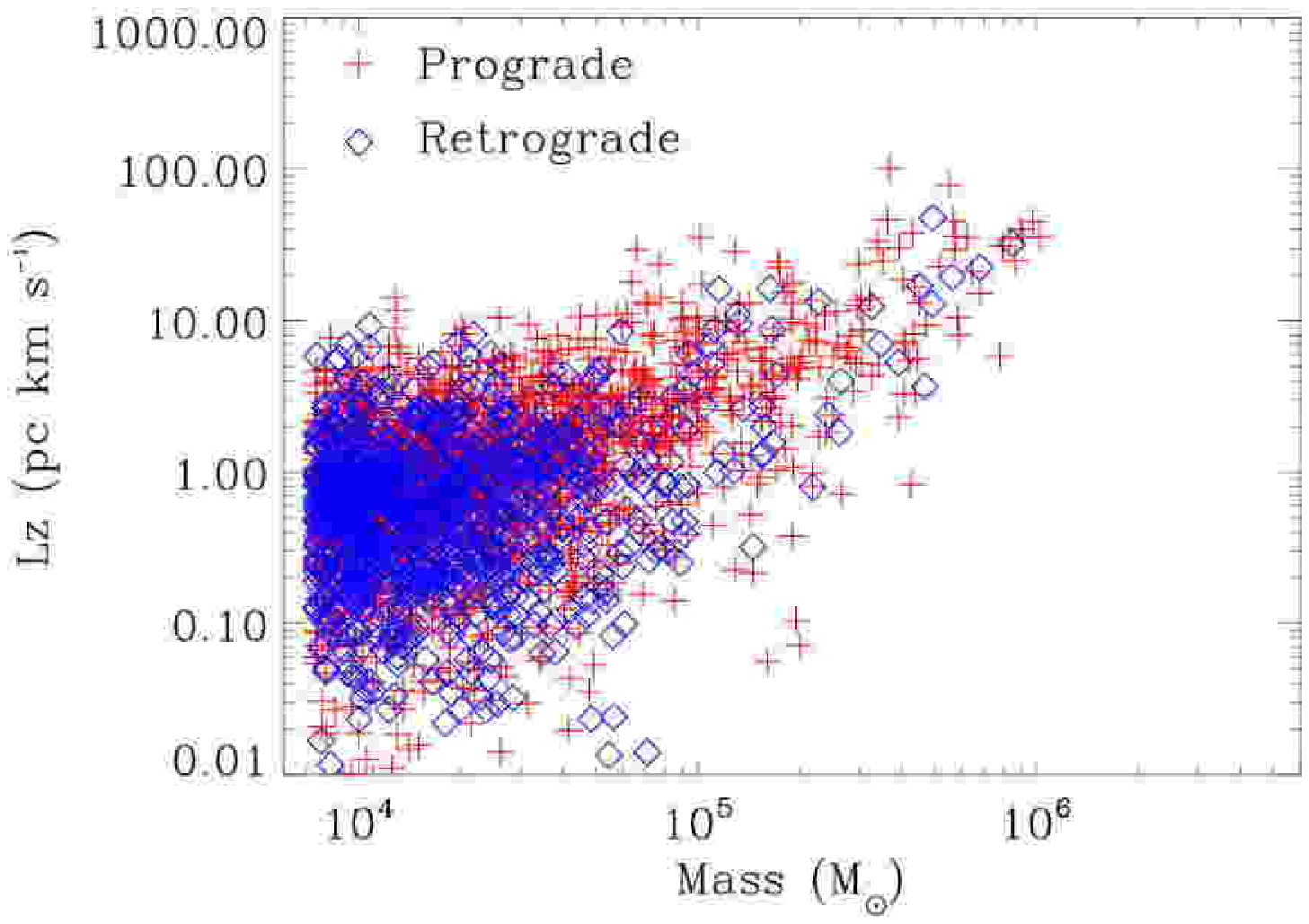}}
\centerline{
\includegraphics[scale=0.55,bb=50 300 500 600,clip=true]{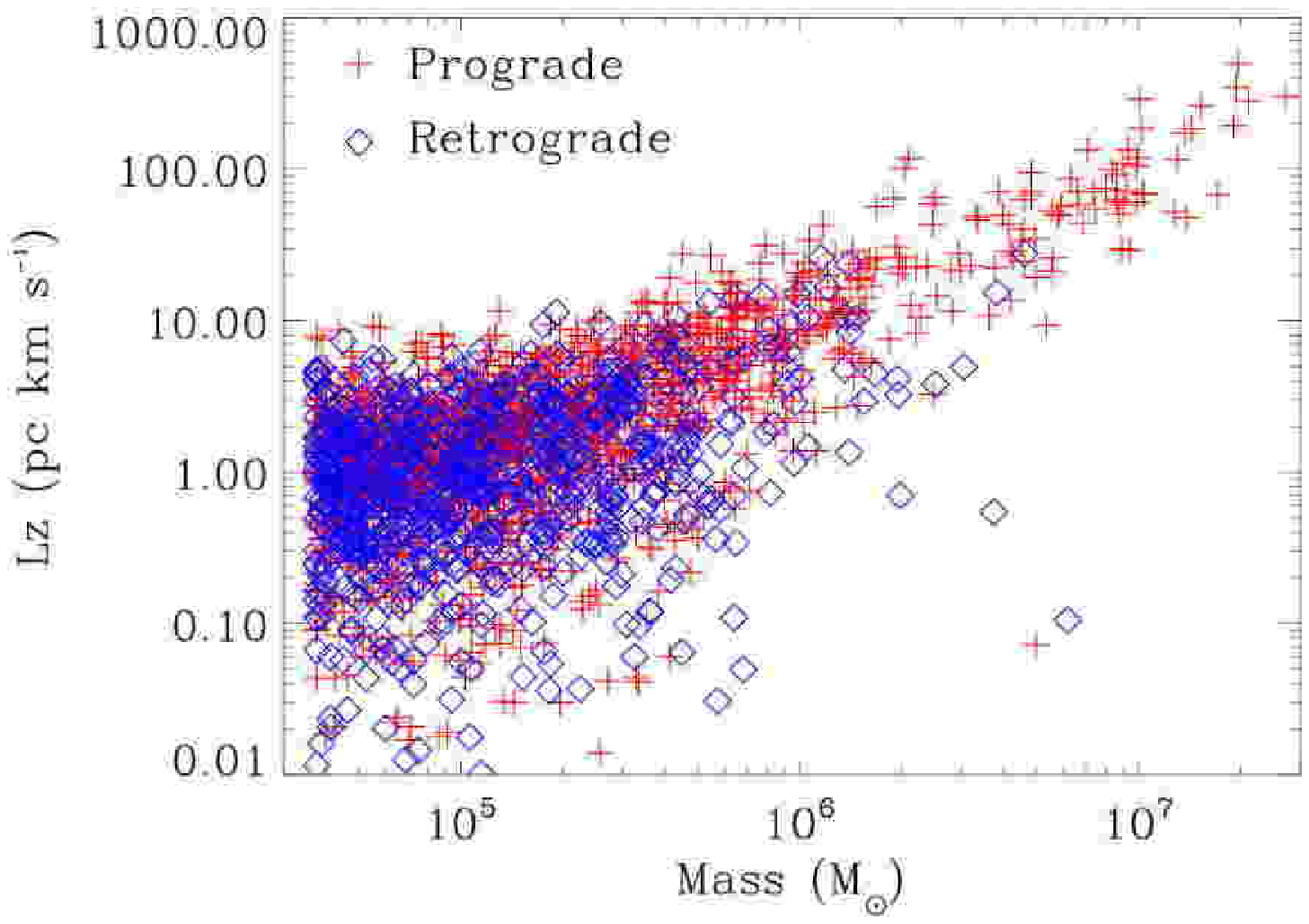}}  
\caption{The specific angular momentum is plotted versus the mass of a cloud for the low surface density case (model B), top, and the high surface density case (model I). Self gravity is included for both. With the higher surface density though, the distribution of angular momentum of the clouds extends to over 100 pc km s$^{-1}$. These high angular momentum clouds are formed mainly by self gravity and exhibit prograde rotation.}
\end{figure} 

\begin{figure}
\centerline{
\includegraphics[scale=0.45]{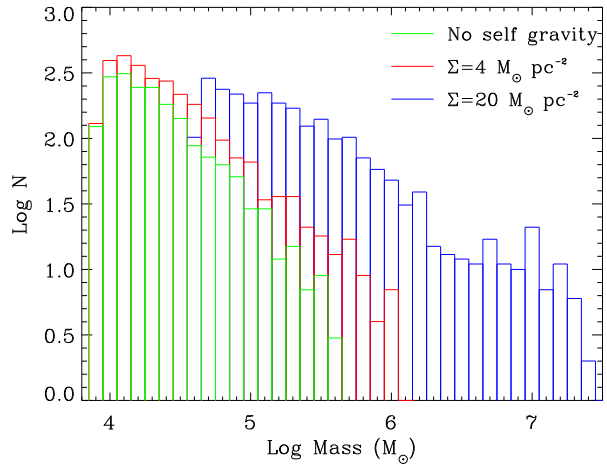}}
\caption{The mass spectrum is shown for clouds extracted from Runs G (red), B (green) and I (blue). The mass spectrum becomes shallower as self gravity becomes more significant. The shallowest slope is obtained when the surface density is 20 M$_{\odot}$ pc$^{-2}$.}
\end{figure} 

\begin{figure*}
\centerline{
\includegraphics[scale=0.35,angle=270]{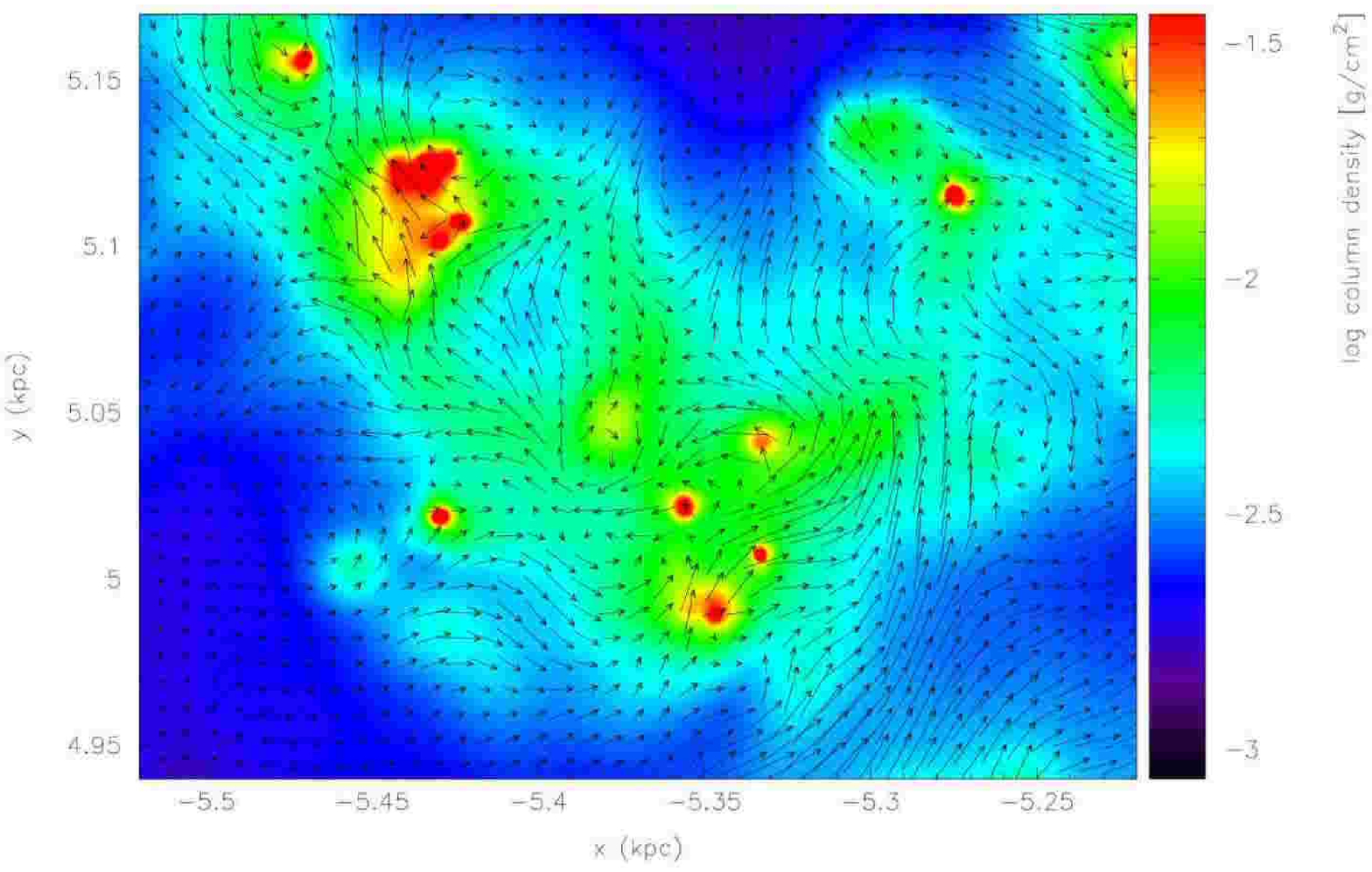}
\includegraphics[scale=0.35,angle=270]{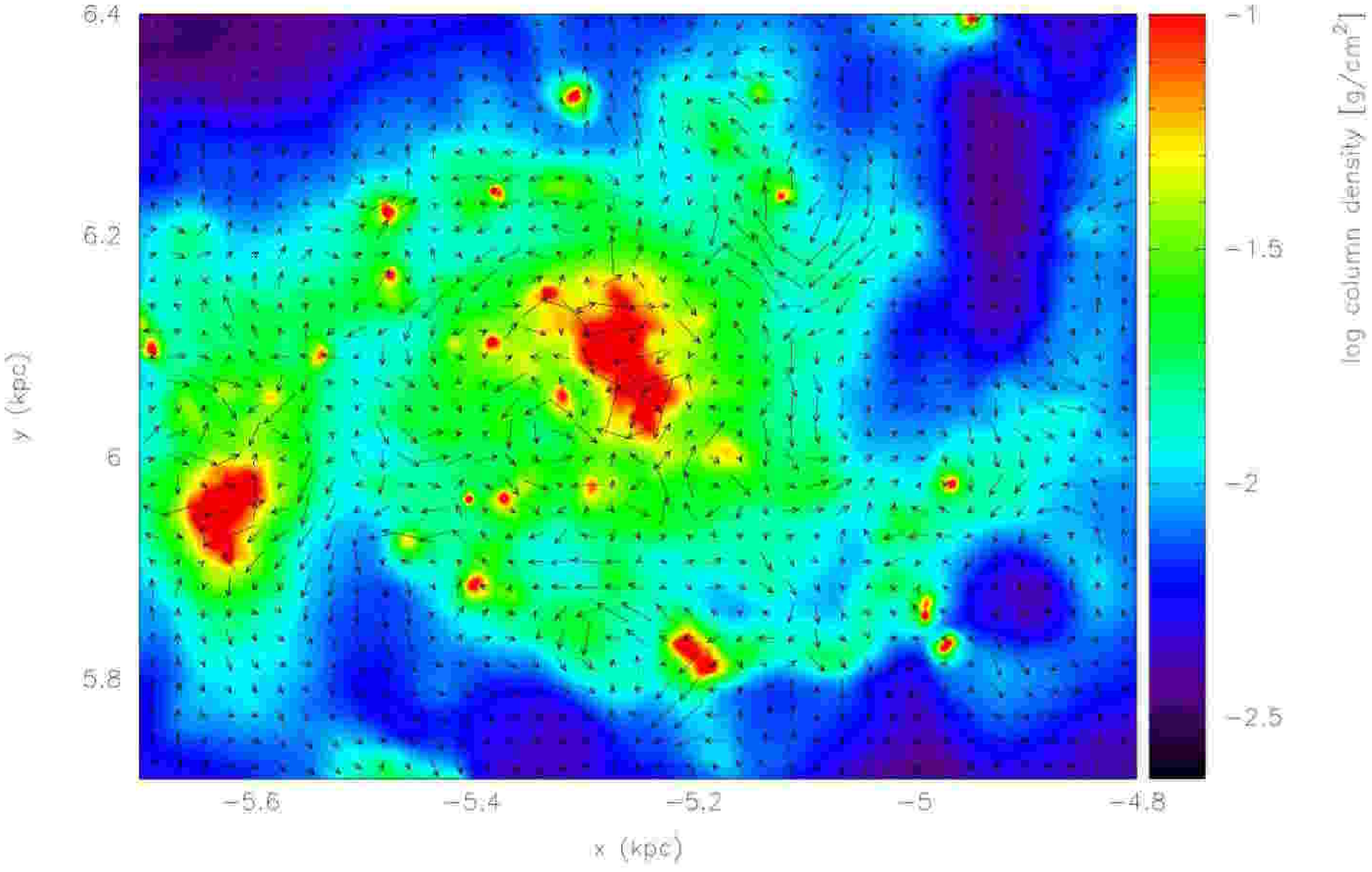}}  
\caption{These figures zoom in on individual structures in the disc which would correspond to GMCs. Vectors show the magnetic field integrated through the plane of the disc. The left panel shows a cloud from the lower surface density model (B), where $\Sigma=4$ M$_{\odot}$ pc$^{-2}$. The mass of the cloud is $3 \times 10^5$ M$_{\odot}$, which is distributed in several separate clumps. The cloud on the right panel contains $\sim 10^7$ M$_{\odot}$ and is taken from model I, with the 20 M$_{\odot}$ pc$^{-2}$ surface density, and self gravity has much more influence on the structure of the cloud. The rotation of the cloud has produced tidal tails extending from the GMC, and the mass of the cloud is concentrated in the centre of the cloud.}
\end{figure*} 
\section{Discussion}
The formation of molecular clouds in spiral arms is believed to occur primarily by gravitational instabilities or cloud collisions. 
For two decades however, agglomeration models for molecular cloud formation have been largely neglected. We have improved previous computational analysis of cloud formation via collisions (e.g. \citealt{Casoli1982,Tomisaka1984,Hausman1984,Kwan1987}) by being able to perform full hydrodynamical simulations, which do not require clump masses or the degree of fragmentation or coalescence as input parameters. 
Furthermore, in our models, the constituent clumps which form the GMCs are generally not self gravitating, avoiding the timescale problems associated with previous models. However, magnetic pressure is also important in the absence of feedback, especially as more massive clouds develop in the spiral arms.

In the lower surface density results we have presented, the formation of GMCs is due to orbit crowding in the spiral shock, which leads to the agglomeration of clumpy gas into larger structures. Without the spiral shock, interactions between clumps are much less frequent and collisions predominantly lead to fragmentation \citep{Gittins2003}. Although fragmentation occurs in our models, the clumps in the spiral arms tend to be more massive and dense than the material entering the shock, so fragmentation is reduced. Orbit crowding was previously suggested by \citet{Roberts1987} as the dominant means of GMC formation in spiral galaxies. Their calculations allowed dissipative collisions between clouds, rather than assuming that colliding clouds coalesce into coherent objects, as was the case for other models at the time. This condition allowed structure to emerge along the spiral arms, similar to our hydrodynamical models. We also find that GMCs formed by orbit crowding tend to be associations of smaller clouds rather than gravitationally bound structures, again in agreement with arguments presented in \citet{Roberts1987}.

Stability analysis of the ISM subject to gravitational collapse has been successful in explaining the observed separation and masses of GMCs along spiral arms. No such analogous interpretation has previously been identified for collisional models. The Jeans analysis however does not explain the spacing between spiral arm clouds in our lower surface density simulations. Instead we interpret this spacing in terms of orbits of the gas and relate the distance between GMCs to the epicyclic radius. We obtain separations of approximately 1-1.6 kpc (although dependent on the parameters for the spiral potential), which is at the lower end of those observed by \citet{Elmegreen1983}, but \citet{LaVigne2006} find spacing between feathers somewhat smaller than this (200 pc -1 kpc depending on galactic radius). For the agglomeration scenario, the mass of GMCs increases for stronger shocks or a higher surface density. Although observations of molecular cloud masses are limited in external galaxies, M51 contains GMCs of $>10^7$ M$_{\odot}$ \citep{Rand1990} whilst galaxies without strong spiral shocks, such as M31 and M33 contain no GMCs of  $>10^6$ M$_{\odot}$ \citep{Sheth2000,Engargiola2003}. However, more massive clouds tend to be formed anyway when the disc is gravitationally unstable.

There are several possible issues with the physics assumed for these calculations.
Although we state the value of $\beta$ in these simulations, in the calculations with cold gas, the magnetic field tends to be lower than observations, as described in \citet{Dobbs2008}. This is because we assume an initially uniform density distribution, whereas the cold gas in the ISM is situated in dense clumps. However, this is unlikely to effect our conclusions, since the ISM is observed to be clumpy despite higher field strengths, and this is the main requirement for agglomeration to occur. Nevertheless, calculations in which the magnetic field strength is consistent with observations in both the warm and cold component are needed in future work.

An important caveat with our results is that we have not included heating and cooling of the ISM. 
Our models require that cold gas exists prior to entering the spiral shock. Observations show a substantial part of the ISM in the Milky Way is located in cold dense HI regions \citep{Heiles2003}, though it is unclear whether this is ubiquitous through the disc, or just associated with gas that is already in the process of forming molecular clouds. If it is the latter, then GMC formation may occur primarily by thermal and gravitational instabilities, with gas cooling and regions fragmenting as collapse occurs. By including a detailed model of the thermodynamic processes of the ISM in our upcoming calculations, we will be able to address this issue.

Furthermore, we have used the thermal sound speed of the gas, rather than the r.m.s. speed associated with turbulence in the gas for the Jeans analysis of the cold gas. From movies of gas flowing through the spiral arm, it is evident that the gas flow is fairly chaotic, and thus the velocity dispersion is somewhat higher than the thermal sound speed of the cold gas. Nonetheless we do not get kpc size clumps supported by turbulence. Our previous simulations of star forming cores also showed that turbulence does not act as an isotropic pressure supporting the cloud against gravitational collapse \citep{Dobbs2005}.  Lastly, we do not include stellar feedback, which will disrupt structure along the spiral arms \citep{Wada2008,Shetty2008}.  

In addition to this further physics, direct comparisons with actual galaxies will be required to evaluate which are the dominant mechanisms for the formation of GMCs. Interestingly, the surface density of our Galaxy is around 10~M$_{\odot}$ within a radius of 12 kpc \citep{Wolfire2003}, roughly at the point gravitational instabilities start to influence the large scale structure in our models.

\section{Conclusion}
We have investigated GMC formation by agglomeration and self gravity. Agglomeration occurs in spiral shocks providing there is a clumpy constituent of the ISM, assumed to be cold HI or molecular gas. This process occurs regardless of whether self gravity is included, and in low surface density calculations, the formation of GMCs is predominantly due to agglomeration. The converse situation arises when there is only warm gas, and GMC formation occurs by gravitational instabilities, providing the disc is gravitationally unstable. Generally, both agglomeration and self gravity are expected to contribute to GMC formation, with self gravity becoming more important to GMC formation and disc structure as the surface density increases.

The degree to which these processes determine GMC properties will depend on the surface density of the galaxy, the thermal nature of the ISM, and most likely the magnetic field strength. In particular, GMCs with retrograde rotation can be produced when the ISM is clumpy. A caveat to these calculations is that we assume a two fluid model with no heating or cooling. We expect to include the thermodynamics of the ISM in future work. A further caveat is that gravitational collapse should occur in our models and lead to stellar feedback. Cooling of the gas to temperatures $<$ 100~K, the inclusion of non-ideal MHD processes, and possibly higher numerical resolution would induce collapse. However the aim of the current paper is to investigate the structure of the disc without feedback. The aim of future simulations will be to see how this picture changes with stellar feedback.

\section*{Acknowledgments}
I am very grateful to Ian Bonnell, Matthew Bate and Jim Pringle for reading a draft version of this paper, and to Daniel Price for useful discussions. I also thank the referee for helpful comments and highlighting issues which required more explanation. Lastly I thank Dave Acreman for maintenance of the Exeter SGI supercomputer, and Chris Rudge for his assistance on UKAFF. 
This work, conducted as part of the award `The formation of stars and planets: Radiation hydrodynamical and magnetohydrodynamical simulations' made under the European Heads of Research Councils and European Science Foundation EURYI (European Young Investigator) Awards scheme, was supported by funds from the Participating Organisations of EURYI and the EC Sixth Framework Programme. 
Computations included in this paper were performed using the UK Astrophysical
Fluids Facility (UKAFF) and Exeter's Astrophysics Group SGI Altix ICE supercomputer, Zen. Figures were produced using SPLASH, a visualisation package for SPH that is publicly available from http://www.astro.ex.ac.uk/people/dprice/splash/  \citep{splash2007}. 
\appendix

\section{Clumpiness of the gas prior to a shock}
\begin{figure}
\centerline{
\includegraphics[scale=0.4]{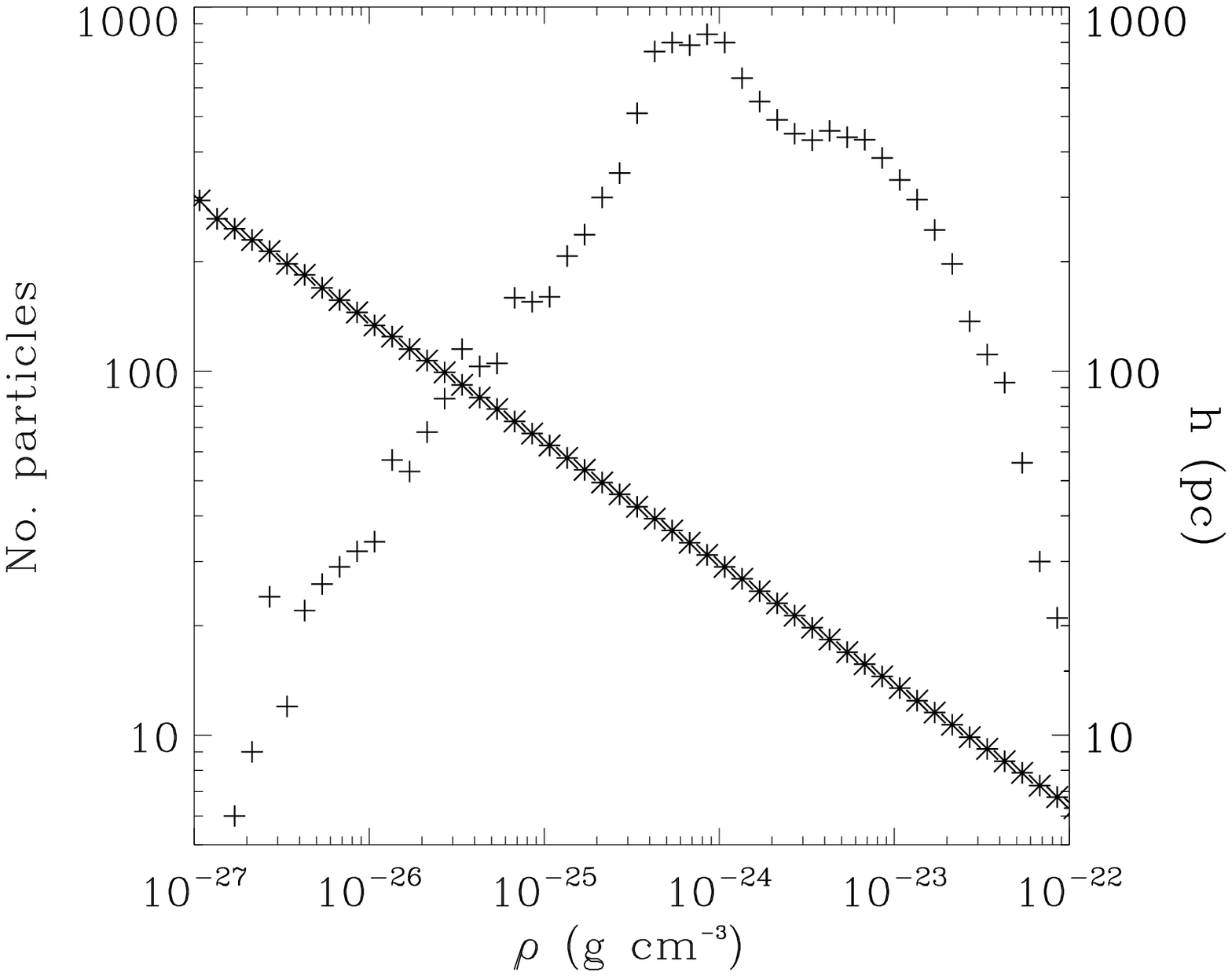}}
\centerline{
\includegraphics[scale=0.4]{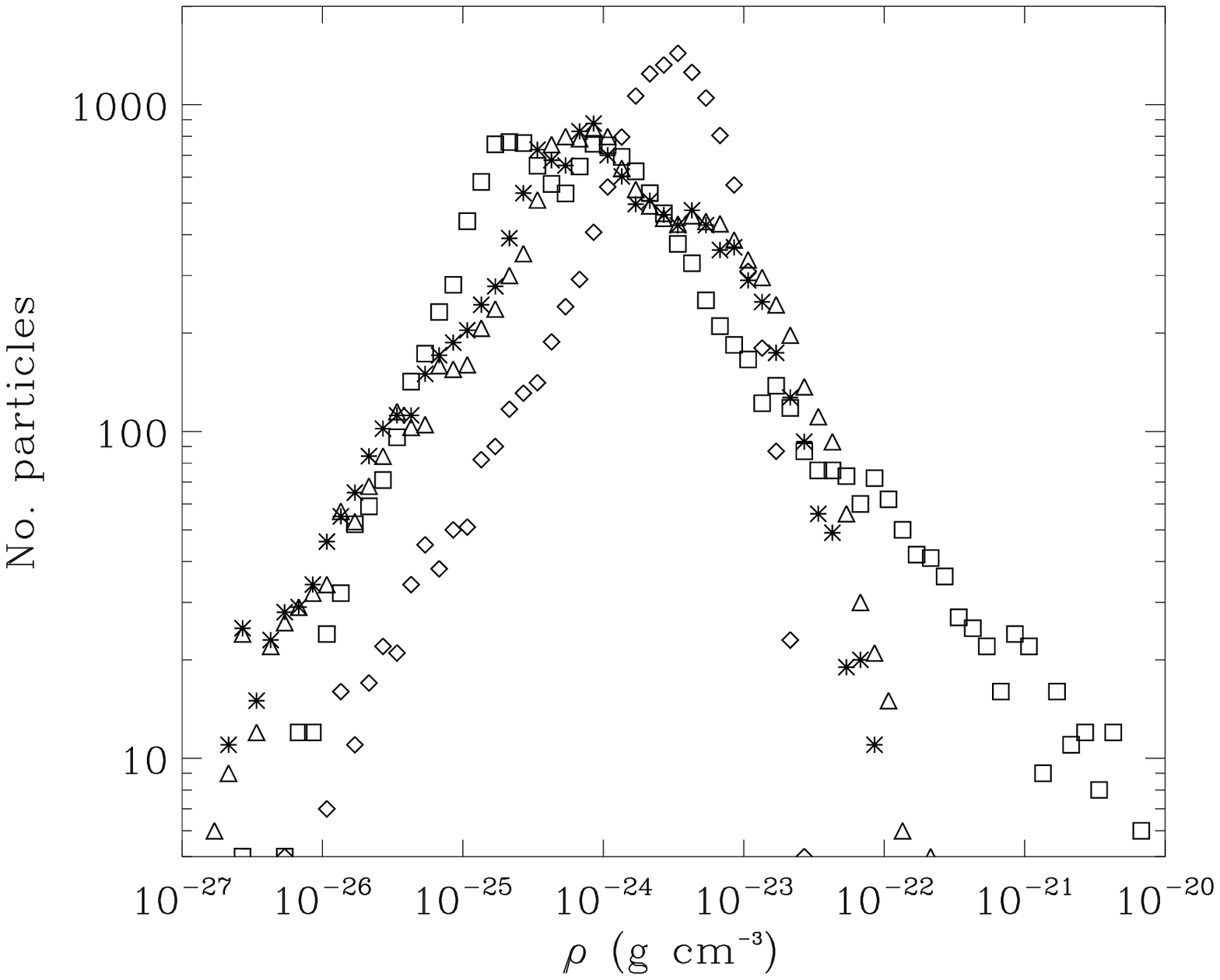}}  
\caption{These panels consider the density distribution of the gas just before entering a spiral shock. Particles are selected from a 1.5 x 1.5 kpc interarm region after 40 and 265 Myr during models B, G and J. The top panel shows the number of particles at different densities (crosses) in model B, the two fluid case with self gravity. The smoothing length (stars) gives an indication of the length scale associated with the structures. The lower panel shows the distribution of particles in the same region for models B (two fluid case with self gravity, triangles), G (two fluid case but no self gravity, stars) and J (100 K gas only, diamonds). There is little difference with or without self gravity. Finally the density distribution is shown at a later time of 265 Myr for model B (squares). The gas has reached much higher densities at this point due to self gravity acting on dense gas in the spiral shock.}
\end{figure} 
In order for agglomeration to occur in the shock, the ISM is assumed to be clumpy. The ISM is observed to be inhomogeneous over many scales, and CNM (cold neutral medium) features are observed in 100 pc \citep{Strasser2007} and parsec/subparsec structures (e.g. \citealt{Dickey1990}). The clumpiness of the gas prior to the shock in our models arises through the Poisson noise of the particle distribution. We show the density distribution of particles before gas has entered the spiral shock in Fig.~A1 (top), for model B. The particles are selected from a 1.5 by 1.5 kpc region after 40 Myr. The particles are about to enter a spiral shock, but at this early stage in the simulation, have not been significantly perturbed by any previous shock. At the very beginning (t=0) there is only an order of magnitude variation in density. However the gas is then subject to the disc potential (Eqn (1)), hence the density varies with scale height  with low density gas at high $z$. Fig.~A1 also shows the smoothing length, $h$, of the particles. So we see that the density spans nearly 5 orders of magnitude, with corresponding length scales of a few to $>$100 pc. 

The density distribution is also shown for different models in Fig.~A1 (lower). The density distribution is narrower for the single phase case, without the warm, lower density, component. With self gravity, the densities are only slightly higher indicating self gravity has little effect on the distribution (self gravity only effects a small fraction of the mass which occupies high densities) before the gas enters the spiral shock. Overall any differences in the distributions for models B, G and J have little impact on the large scale structure that emerges because the size scales of the initial structure is always much less than the sizes of the large clouds that emerge later in the simulations. Finally in Fig.~A1, the density distribution is shown for an interarm region at a much later time in model B. The much higher densities indicate the influence of self gravity in the spiral shock. The density of the gas increases in the shock such that self gravity has much more of an effect. Hence clumps with much higher internal densities occur at later times in the simulation (and compared to the non-self gravitating model).          
\section{Resolution study}
\begin{figure*}
\centerline{
\includegraphics[scale=0.52,bb=0 20 550 450,clip=true]{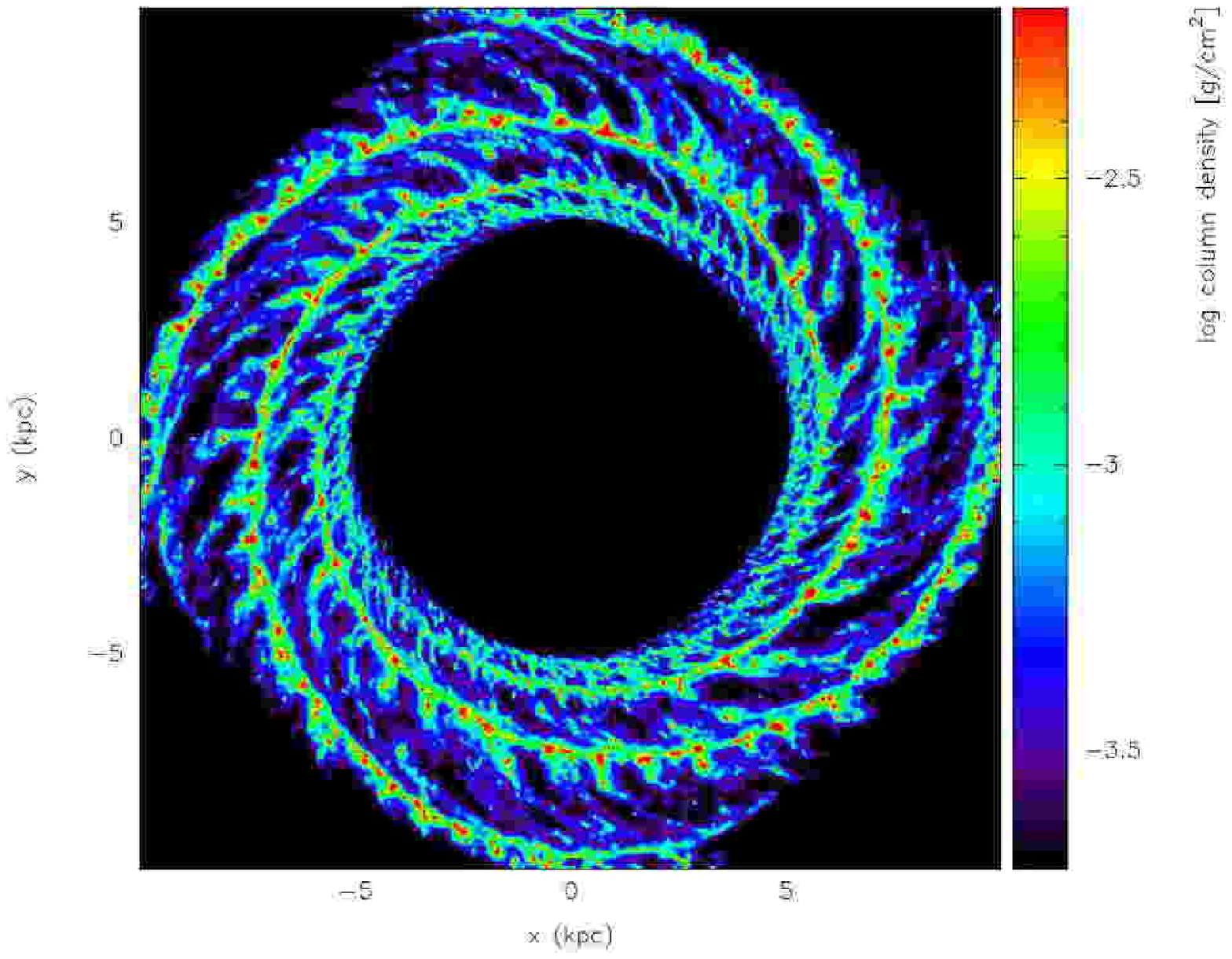}
\includegraphics[scale=0.52,bb=50 20 550 450,clip=true]{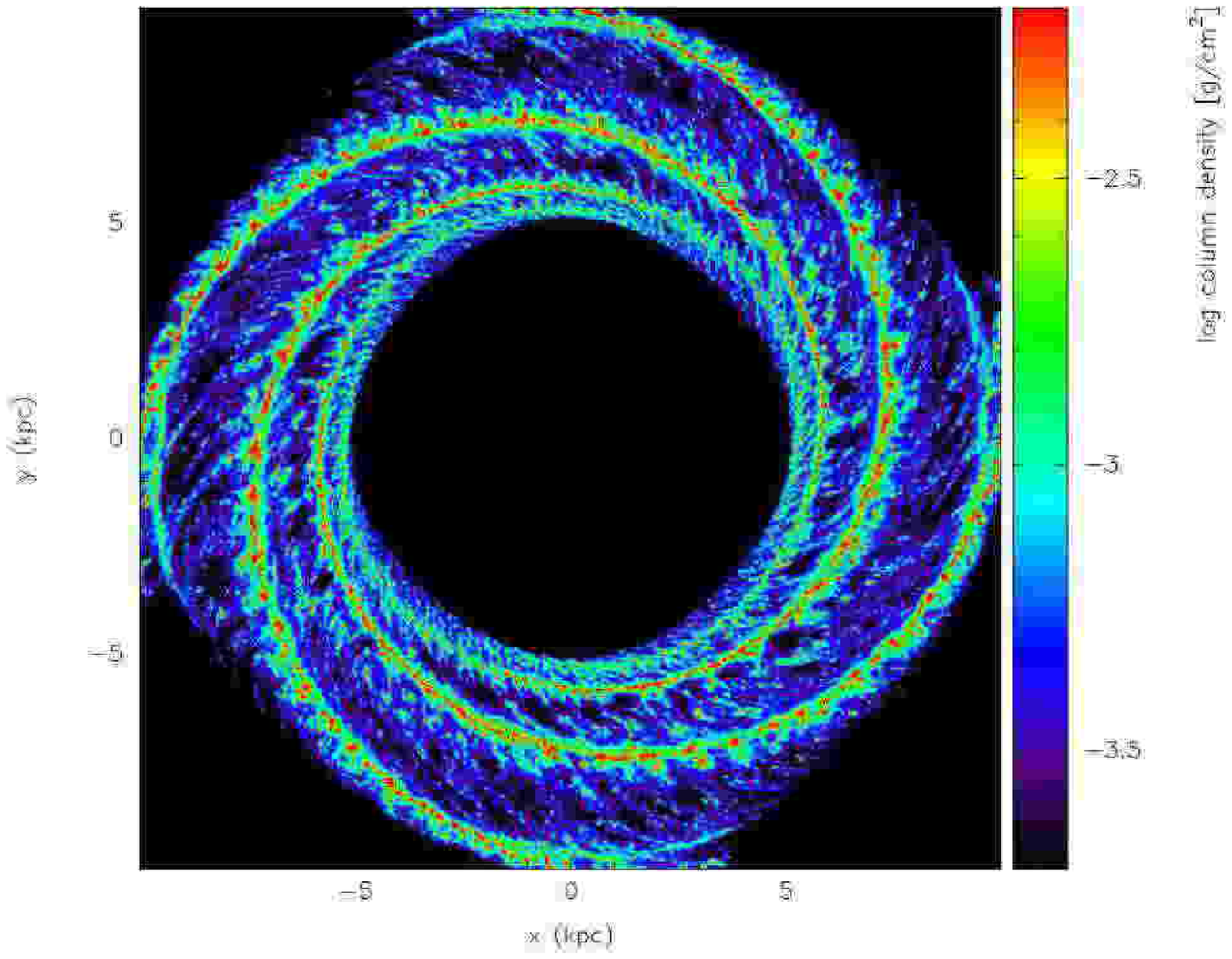}} 
\centerline{
\includegraphics[scale=0.52,bb=0 20 550 400,clip=true]{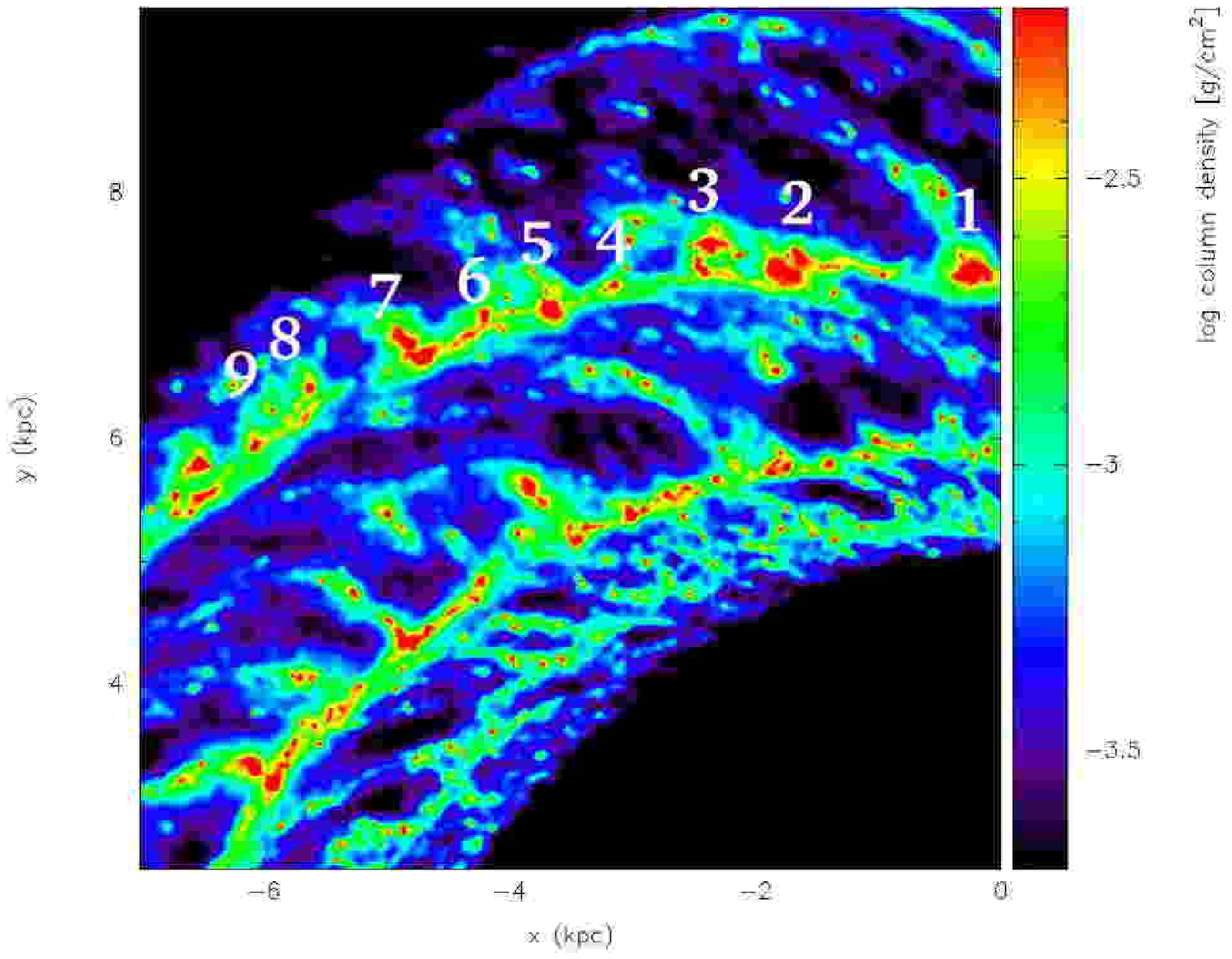}
\includegraphics[scale=0.52,bb=50 20 550 400,clip=true]{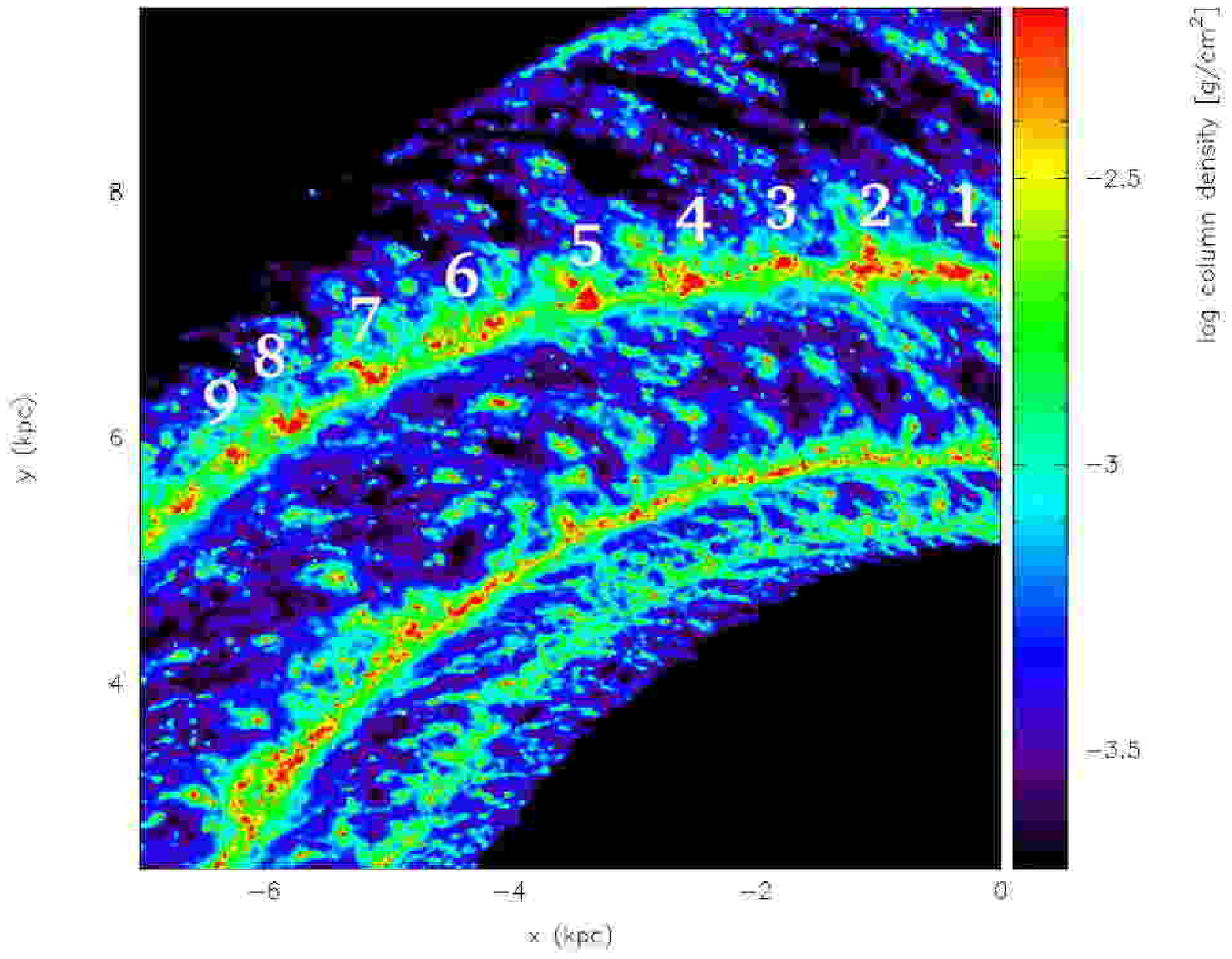}}
\caption{This figure shows the disc and a subsection of the disc from model J, with 1 million particles (left) and 8 million particles (right). The equivalent simulation with 4 million particles is shown in Fig.~1d). The lower panels intend to show that the location of large scale features is similar in each simulation and in particular, the number of large clumps per unit length along the spiral arm does not change. However there is clearly much more substructure in the higher resolution calculations and interarm spurs are less well defined. The inner arm appears more continuous because the spiral potential is weaker at these radii (the strength of the spiral potential from \citet{Cox2002} has a maximum at R=8 kpc).}
\end{figure*}

We investigate the difference in structure depending on resolution by repeating model J with 1 and 8 million particles. The results are shown in Fig.~B1, which displays the whole, and a subsection of the disc. The structure of the disc clearly changes with resolution. The interarm spurs are less distinct and more disjointed in the 8 compared to 4 million particle simulation. However the actual location and number of features along the spiral arm appear similar at different resolutions. In particular the number of large clumps per unit length is unchanged. Although some of the spiral arm clumps are much easier to distinguish than others: structures such as that labelled `6' are instead collections of many smaller clumps. The similarity of spacing between the clumps is expected, since the size scale of the initial clumpiness in the gas is much smaller than the resulting size scale of clouds and spurs which form in the disc. However for higher resolution, it is more difficult to assemble gas into a coherent, single cloud.

Overall, the increased resolution has a similar effect to a previous resolution comparison in \citet{Dobbs2008}, in that the same large scale features occur in different resolution simulations, but the structure is more disjointed at higher resolution. In the calculations here, the gas also reaches higher densities in the 8 million particle case (compared to 1 million), since the dense regions of GMCs are better resolved.
\bibliographystyle{mn2e}
\bibliography{Dobbs}

\bsp

\label{lastpage}

\end{document}